\documentclass[a4paper,12pt]{article}
\usepackage{amssymb}

\usepackage{amsmath}
\usepackage{graphicx}
\usepackage{tabulary}
\usepackage{multirow}
\usepackage[table,xcdraw]{xcolor}

\newcommand{\beq}{\begin{equation}}
\newcommand{\eeq}{\end{equation}}

\newcommand{\Rset}{{\mathbb R}}
\newcommand{\eq}[1]{(\ref{#1})}
\newcommand{\pain}{Painlev\'{e} }
\newcommand{\painx}{Painlev\'{e}}
\newcommand{\mumin}{\mu_{P,\rm{min}}}
\begin{document}

\title{The \pain paradox in contact mechanics}

\author{Alan R. Champneys, P\'eter L. V\'arkonyi}
\date{final version 11th Jan 2015}
\maketitle

\begin{abstract}
The 120-year old so-called Painlev\'{e} paradox involves the loss of
determinism in models of planar rigid bodies in point contact with a
rigid surface, subject to Coulomb-like dry friction.  The phenomenon
occurs due to coupling between normal and rotational
degrees-of-freedom such that the effective normal force becomes
attractive rather than repulsive. Despite a rich literature, the
forward evolution problem remains unsolved other than in certain
restricted cases in 2D with single contact points.  Various practical
consequences of the theory are revisited, including models for robotic
manipulators, and the strange behaviour of chalk when pushed rather
than dragged across a blackboard.

Reviewing recent theory, a general formulation is proposed, including
a Poisson or energetic impact law. The general problem in 2D with a
single point of contact is discussed and cases or inconsistency or
indeterminacy enumerated. Strategies to resolve the paradox via
contact regularisation are discussed from a dynamical systems point of
view. By passing to the infinite stiffness limit and allowing impact
without collision, inconsistent and indeterminate cases are shown to
be resolvable for all open sets of conditions. However, two
unavoidable ambiguities that can be reached in finite time are
discussed in detail, so called dynamic jam and reverse chatter. A
partial review is given of 2D cases with two points of contact showing
how a greater complexity of inconsistency and indeterminacy can
arise. Extension to fully three-dimensional analysis is briefly
considered and shown to lead to further possible singularities.  In
conclusion, the ubiquity of the \pain paradox is highlighted and open
problems are discussed.
\end{abstract}

{\bf KEYWORDS: \textit{rigid body}; \textit{Painlev\'{e} paradox}; \textit{Coulomb friction}; \textit{impact mechanics}; 
\textit{dynamic jam}; \textit{sprag-slip oscillation}; \textit{chatter}; 
\textit{Zeno phenomenon}; 
\textit{reverse chatter}; 
\textit{impact without collision}; \textit{inconsistency};
\textit{indeterminacy}; \textit{rational mechanics}.}

\section{Introduction}

Paul Painlev\'{e} is a truly remarkable figure in the history of
science and technology. In mathematics, Painlev\'{e} is most closely
associated with the family of integrable differential equations and
corresponding transcendental functions that bear his name.  But there
is much more to Painlev\'{e}, \cite{Borisov}; he made pivotal
contributions to the theory of gravitation, he was possibly the
first ever aircraft passenger (taking a seat in the Wright brother's plane in
1908) and set up the world's first degree in programme in Aerospace
Engineering. His role as Minister of Public Instruction and
Inventions, led to him being appointed Prime Minister of France
briefly during the First World War, and again in 1924. In the latter
stint he was forced to resign after being unable to solve the
overriding financial crisis of the time (an uncanny parallel to 21st Century
European politics, perhaps).

This paper shall look at the legacy of Painlev\'{e}'s work in
rigid body mechanics, specifically the rather curious phenomena
that are present in models of dry friction.  In a series
of papers starting in 1895
\cite{Painleve1895,Painleve1905a,Painleve1905b}, \pain provided
simple examples which illustrate a fundamental inconsistency in the
laws of Coulomb friction. (Technically, we should refer to
Amontons-Coulomb friction, because the laws were first postulated by
Amontons at the beginning of the 18th Century, whereas Coulomb's role
was one of experimental verification, see
e.g.~\cite{Martyn1742,Feeny1998}. As with many unjust attributions in 
science, Amontons' name seems to have been dropped from the epithet now used 
to describe the simplest law of dry friction. In fact, theory that 
additionally involves solving for
the normal force at contact is sometimes called the Signorini-Coulomb model 
or the Signorini-Amontons-Coulomb $\ldots$, but we shall just refer to 
Coulomb friction in what follows.) 
The so-called Painlev\'{e} paradox (which, to add more historical accuracy,
as Painlev\'{e} acknowledges, was first discovered by Jellet \cite{Jellet1872})
occurs in planar bodies undergoing oblique contact for which
there is a sufficiently large coupling between normal and tangential
degrees of freedom at the contact point. Such configurations, given
sufficiently high coefficient of friction, can reach a point where
there is no longer a unique forward simulation --- either a multiplicity of
possibilities (\textit{indeterminacy}), or none (\textit{inconsistency}) --- so there
is no way to decide what happens next within rigid body mechanics. 
Further historical remarks on the origins of the Painlev\'{e} paradox
can be found in the books by Brogliato \cite{Brogliato1999} and
and Le Suan Anh \cite{LeSuanAn_book}.

\begin{figure}
\begin{center}
\includegraphics[width=0.35\textwidth]{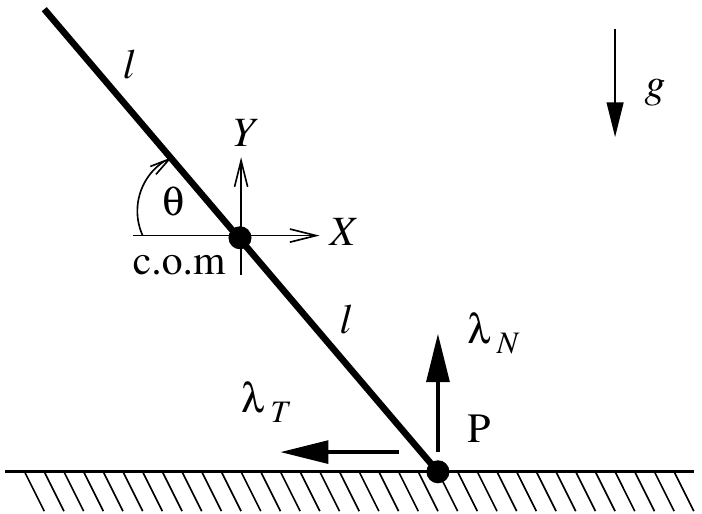}
\end{center}
\caption{The classical \pain problem (CPP) for a falling rod.}
\label{fig:CPP}
\end{figure}

The configuration most closely associated with Painlev\'{e} is that of
rod falling under gravity whose lowest end is in contact with a
horizontal surface and subject to simple Coulomb friction, see
Fig.~\ref{fig:CPP}.  We shall refer to this as the classical \pain
problem (CPP), although, as pointed out by Leine \cite{Leine2002}, this was
not actually the example that Painlev\'{e} considered first (see
Sec.~\ref{sec:2}).

The CPP can be realised using a pencil that is momentarily balanced on
its tip on a rough piece of paper attached to a rigid horizontal
table.  Upon letting go, minor disturbances mean the pencil will have
a shallow angle to the vertical and will start to fall.  Initially,
the tip remains fixed, forming a pivot about which the pencil
rotates. Upon reaching some critical angle, the tangential force at the
tip will be large enough for the pencil to begin to slip (in the
opposite direction to the instantaneous horizontal velocity component
of the top). A faint line begins to appear on the
paper. As the pencil continues to fall, 
before it becomes fully horizontal, the angular acceleration will be
sufficient to overcome the vertical component of the ground reaction
force at the tip.  Thus, the tip lifts off and the faint line is
terminated. A fraction of a second later, the blunt end of the
pencil hits the table, some rattling motion ensues, before the pencil
becomes fully horizontal, rolls for a while and comes to rest on the
table. No paradox.

What \pain realised though is that if the coefficient of friction is
large enough (in fact, unrealistically large for an approximately
uniform rod like a pencil on anything other than the coarsest
sandpaper) then something else can happen. Before the pencil lifts
off, it can reach a configuration that is now known as \textit{dynamic
  jam} where one cannot decide with certainty what happens next.  The
pencil tip could lift off in the usual way with an
initially infinitesimal normal velocity.  Or, the pencil could remain in contact
such that the more
it tries to lift off, the more it rotates and digs back down into the surface. 
It is as if there is a negative normal force pulling the tip
into the surface. If one adds a rigid impact model,
like Newton's law of restitution, 
to the mechanical
formulation, 
then such a
force causes a so-called \textit{impact without collision} (IWC)
\cite[Ch.~5.5]{Brogliato1999}, see Sec.~\ref{sec:3.2} below. Such
an impact involves an impulsive jump to reduce the tip's slipping
velocity to zero and eject it from the surface with
a finite normal velocity. We give a more quantitative
description of the paradox for the CPP in Sec.~\ref{sec:2.2} below.

\begin{figure}
\begin{center}
\includegraphics[width=0.7\textwidth]{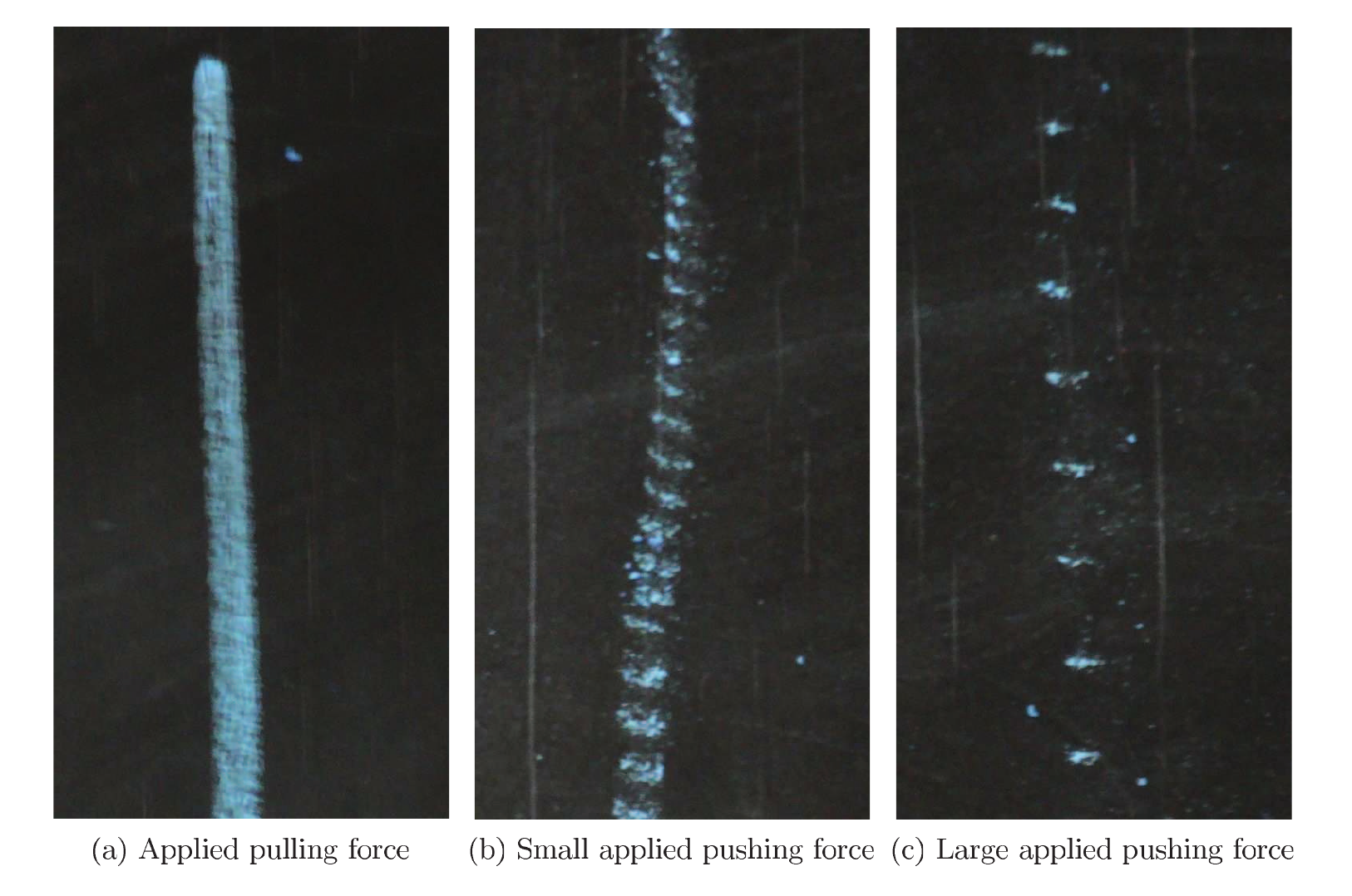}
\label{fig:hall}
\end{center}
\caption{After \cite{Hall}, previously unpublished, reproduced by permission.
Three traces left by chalk moved by hand 
across a rigid blackboard. In all figures the
direction of motion is downwards. In (a) there is an acute angle between
the chalk's axis and the velocity vector, whereas in (b) and (c) the angle
is obtuse. The difference between (b) and (c) is that an increased
normal force is applied in (c).} 
\end{figure}

One of the purposes of this paper is to show that the \pain paradox is
\textit{not} just a theoretical curiosity manifest only in ``toy''
mathematical models of pencils on unrealistically rough surfaces.  The
consequences of the Painlev\'{e} paradox are in fact ubiquitous, even
in everyday phenomena. For example, robotic manipulators, pieces of
chalk or even your finger are known to \textit{judder} when they are being
pushed across a rough surface \cite{Liu2007}, see
Sec.~\ref{sec:2.3} for more details.  The phenomenon in question,
sometimes referred to \textit{sprag-slip oscillation} \cite{Hoffmann}, is
more complex than mere stick-slip behaviour, as it additionally
involves lift-off and impact every cycle.  In fact, the somewhat
controversial American physicist Walter Lewin is well known for
demonstrating how to exploit this effect to draw 
dotted lines on a blackboard reliably and rapidly
by inducing sprag-slip oscillations 
in the chalk. See Fig.~\ref{fig:hall}

Nordmark \textit{et al} \cite{paper2} demonstrated that such
sprag-slip oscillations can arise as a consequence of the \pain paradox through
an instability that they termed \textit{reverse chatter} (see Sec. \ref{sec:3}
below). This instability involves a sequence of impacts that
accumulate backwards in time; rather like watching a video in reverse 
of a bouncing ball coming to rest.  Another everyday phenomenon
that can be considered to be a consequence of the \pain paradox is how
a rigid body can seem to defy gravity by being ``wedged'' into an
overhang (see Fig.~\ref{fig:ambiguity} below).
As we shall show in Sec.~\ref{sec:6}, there can be subtle
effects that involve the interplay between the two contact points,
each of which can independently exhibit the \pain paradox.

The main aim of this paper is to provide an overview of recent
academic work relating to the \pain paradox in order to provide some
clarity, while keeping in mind the practical implications of the
theory.  In so doing, we shall also include preliminary new results by
the authors and their collaborators, the details of which shall appear
elsewhere.  Overall, we shall attempt to provide answers to the
following two central questions:
\begin{description}
\item[Question 1.] For a given rigid body system 
with contact, in a given state at a
  given instance of time, is there a uniquely defined forward-time
  continuation of the dynamics within the formalism of rigid body
  mechanics?  This divides naturally into two subquestions: is \textit{any}
  possible forward continuation possible (consistency), 
and, if so, is it unique (determinacy)?  
\item[Question 2.] Given a rigid body system 
that is not exhibiting the \pain
  paradox at one instance of time, what possible ways can it enter
  into a \pain paradox state at a later in time? Again there are two
  subquestions: how unlikely is it to enter an inconsistent or
  indeterminate state; and what happens next?
\end{description}

As we shall show, no complete answer to either question is yet
known. Nevertheless, we shall attempt to provide partial answers,
restricted to the case of a single point contact in two spatial
dimensions. As we shall see though, there are way more possibilities
that need consideration in 3D or in the presence of several isolated
point contacts, for which we can only provide partial results. For
more physically realistic problems with regional (line or patch)
contacts, almost nothing is known. Therefore in this paper we shall
restrict attention to problems with a finite number of \textit{isolated
  point contacts}.

Another aim of this paper is steer a path through the various
approaches that have been proposed to \textit{resolve} the \pain paradox
from both a practical and a rational, analytic point of view.  
Our focus is not particularly on numerical methods for simulation in the presence
of the paradox, although it is worth pointing out recent results for
example in \cite{Acary,Schindler,Trinkle}. Nor do we consider 
control algorithms designed to avoid \pain paradoxes in robotic systems, but
see e.g.~\cite{Brogliato1999,Elkar:11} for recent results. 
Instead, we shall 
attempt to understand the dynamical consequences of the \pain paradox
from a mathematical modelling point of view, 
building on recent
understanding of the theory of piecewise-smooth
dynamical systems
see \cite{Filippov,Leine_book,SIAM_review,Springerbook},
In fact, it is well known that piecewise-smooth systems of
Filippov-type \cite{Filippov} can exhibit nonuniqueness or nonexistence of solutions, see e.g.~\cite{Filippov,Bastien,Jeffrey}.  

Philosophically speaking, the \pain paradox is not a puzzle about
the real world, but a failure of a theory based on 
rigid body mechanics and Coulomb friction to provide complete unequivocal
descriptions of dynamics. 
In truth, any physical resolution of the \pain paradox tends to involve 
relaxing some of
the assumptions of rigid body mechanics in order to predict what
happens next, see Sec.~\ref{sec:4}. 
In effect, the \pain paradox provides insight into
points of extreme sensitivity in rigid body mechanics where additional
physics, possibly at the microscale, is required in order to
accurately capture the dynamics.  
Such points of extreme sensitivity
are typically going to be observable as points where significant changes
can result from minuscule variations of the physical properties or 
initial states of the bodies involved. A key question to be addressed then is:
\begin{description}
\item[Question 3.]  Given a rigid body system with contact whose forward
  evolution is either inconsistent or indeterminate, what is the
  minimal extra modelling ingredient in order for there to be a unique
  forward evolution in all inconsistent and indeterminate cases?
  Which cases can be made \textit{uniformly resolvable} via this
  approach; by which we mean, if the additional ingredient to the
  theory arises through a small parameter $\varepsilon$, does a
  qualitatively consistent outcome occur uniformly in the limit as
  $\varepsilon \to 0$?
\end{description}

The rest of this paper is outlined as follows. Section \ref{sec:2}
contains a historical introduction to the subject. A mathematical
explanation of the paradox is given in terms of the CPP. This is
followed by a discussion of the role of different friction laws, a
description of other simple configurations that feature
the \pain paradox and a brief review of the various attempts
to resolve the paradox.  Section \ref{sec:3} then introduces a general
formulation for the simplified case of a planar system with a
single contact. An attempt is made 
to answer Question 1 by enumerating all the different cases that lead
to inconsistency or indeterminacy.  We also discuss how to augment the
formulation by including an impact law, and discuss the possibility of
the accumulation of impacts via \textit{chatter}
(also known as the Zeno phenomenon).  Section \ref{sec:4} then 
looks at Question 3 by adding compliance
into the model in order to find cases that are uniformly resolvable.
Section \ref{sec:5} considers Question 2 for the 2D single contact
case; the theory of G\'{e}not and Brogliato \cite{Genot1999} is
reviewed and generalised, to show cases where dynamic jam must
occur. We also consider whether the \pain paradox can be 
approached through chatter and consider conditions under which 
reverse chatter can be triggered. 
Section \ref{sec:6} considers extensions to the theory for configurations with
two or more contact points and starts to
enumerate the additional paradoxes that can occur.
Section \ref{sec:7} then briefly considers the single-contact problem in 3D, 
illustrating that a different kind of dynamic jam can occur which does
not involve the singularity analysed in \cite{Genot1999}.
Finally, Section \ref{sec:8} draws conclusions, highlights 
open problems and and makes philosophical remarks.

\section{Historical perspective}
\label{sec:2}

\subsection{The classical \pain problem}

The CPP, first proposed by \pain in 1905
\cite{Painleve1905b}, involves a slender rigid rod falling under
gravity while in contact with a rigid, stationary, frictional
horizontal surface. Suppose the rod has mass $m$ and radius of
gyration $r$, that the centre of mass is a distance $\ell$ from the
contact point $P$, and that $\lambda_N \geq 0$ and $\lambda_T$ are the
normal and tangential components of contact force.  Let $(X,Y)$ be the
Cartesian co-ordinates of the rod's centre of mass within the plane
in which it is falling, and $\theta$ be its angle to the horizontal,
as depicted in Fig.~\ref{fig:CPP} (note that for convenience we have chosen
an angle $\theta$ which is $\pi$ minus the angle of the same name used
in \cite{Genot1999}).  
Then, under the usual assumptions
of Lagrangian mechanics, the equations of motion can be written as
\begin{equation}
m\ddot{X} = -\mu \lambda_T, \qquad m \ddot{Y} = -mg + \lambda_N
\qquad mr^2 \ddot{\theta} = - \ell(\cos \theta \lambda_N + \mu \sin \theta 
\lambda_T).
\label{eq:CPP}
\end{equation}
Let 
$$
(x,y):=(X+\ell \cos \theta, Y-\ell \sin \theta), \qquad \mbox{and} \quad
(u,v):=(\dot{x},\dot{y})
$$
represent the generalised tangential and normal co-ordinates and velocities
at $P$.  Then, $\lambda_N$ in \eqref{eq:CPP} can be considered to be 
a dynamic Lagrange multiplier that is used to maintain the inequality
constraint $y\geq 0$. In particular $\lambda_N$ and $y$ are said to satisfy
a \textit{complementarity relation}
$\lambda_N \perp y >0$, which means that at most one of $\lambda_N$ and
$y$ can be positive (see e.g.~\cite{Brogliato1999}).
Whenever $y=0$, we assume a simple Coulomb friction law:
\beq
|\lambda_T| \leq \mu |\lambda_N|, \qquad 
 \lambda_T =  - \mu \: \mbox{sign}(u) \lambda_N  \quad \mbox{for} \: u \neq 0,
\label{eq:Coulomb}
\eeq
where $\mu$ is the \textit{coefficient of friction}. 

Now, suppose there is an initial condition such that the rod is slipping 
with $u>0$ and 
$0<\theta<\pi/2$, so that $\lambda_T=-\mu \lambda_N$.
Then, using \eqref{eq:CPP}, 
the normal acceleration can be written as
\begin{eqnarray}
\ddot{y} & = & \ddot{Y} + \ell( \dot{\theta}^2 \sin \theta 
- \ddot{\theta} \cos \theta ) \: , \nonumber \\
         & = & (\ell \dot{\theta}^2 \sin \theta - g) + 
\left [1 + \frac{\ell^2}{r^2} \cos^2 \theta - \mu \cos \theta \sin \theta) 
\right ] \frac{\lambda_N}{m} \: , \nonumber \\
       & & \nonumber \\
       & := & b(\theta,\dot{\theta}) \quad + \quad p^+(\theta,\mu) \lambda_N \: .
\label{eq:CPP_bp}
\end{eqnarray}

Equation \eqref{eq:CPP_bp} describes the dynamics of the CPP in the 
normal direction at the contact point, in terms of two scalar quantities
$b$ which describes the free normal acceleration in the absence
of any contact forces and $p^+$ which we shall refer to as 
the ``degree of \pain-ness'' or the 
\textit{Painlev\'{e} parameter}. In fact, the rest of this paper shall make 
extensive use of these two dynamic scalar quantities (and a third, 
$p^-$, the equivalent of $p^+$ for slipping in the other direction) 
for quite general 
classes of contact problems. 
As we shall see in Sec. \ref{sec:2}, 
$b$ will always be a function of generalised velocities and co-ordinates, 
whereas for stationary contact surfaces $p^\pm$ are
functions of position variables and properties of the friction law only.

In the case of the CPP, equation \eq{eq:CPP_bp} shows that if the rod starts at
rest in the near vertical position (with an angle 
$\theta$ just smaller than $\pi/2$), then clearly $b<0$ and $p^+>0$ initially.
Moreover $b$ and $p^+$ are smooth functions of dynamic variables so, as long
as the rod maintains the condition $u>0$ for slip, these quantities 
will evolve smoothly and preserve their signs for small times.
Hence, for sufficiently short times there is a unique normal force 
\beq
\lambda_N=-b/p^+ >0
\label{eq:slip_normal_force}
\eeq
that makes the normal acceleration in \eqref{eq:CPP_bp} 
vanish, so that the rod remains in contact while it falls. 

Similarly if $\theta$ is initially 
small and positive so that the rod is close to horizontal
then $b<0$ and $p^+>0$. However, for intermediate angles, depending on
other parameters and velocities, $p^+$ and $b$ may in general
take either sign. So let us analyse what happens at a time when
either $p$ or $b$ smoothly changes sign during motion.

A point where $b$ passes from negative to positive (while $p^+$ remains positive) 
is easy to analyse. This would represent a point at which the rod 
simply lifts off from the surface. The normal force 
\eqref{eq:slip_normal_force} smoothly tends to zero, 
the dynamics would lift off into free motion with $y(t)>0$ and $\lambda_N=0$.  

The case of negative $p^+$ is much more unusual.  
First, consider the possibility of
$p^+<0$ and $b<0$. Here, the free acceleration $b$ pushes the tip
of the rod down towards the surface. However the normal force given by
\eqref{eq:slip_normal_force} would be negative, which violates our
complementarity assumption. A positive reaction force
$\lambda_N$ would cause acceleration in the same
direction as $b$, pushing the rod down further into the surface, precluding any
vertical equilibrium.  The rod cannot remain in
contact with the surface, ergo it must lift off. But it can't because
if we look for free motion with $\lambda_N=0$, the free acceleration
$b<0$ takes us back into contact.  Thus, we 
have a configuration that is \textit{inconsistent}, and there is no
valid continuation of the motion forwards in time.

Consider instead an initial conditions for which $p^+<0$ and
$b>0$. Here, in the absence of any contact forces, the rod would
simply lift off with $\lambda_N=0$. However there is another
possibility; there is now a unique non-zero normal force $\lambda_N
=-b/p >0$ for which the free normal acceleration $b$ is
equilibrated. So the rod could remain in contact.  Hence, this case is
\textit{indeterminate}, because there is non-uniqueness in possible
outcome.

It is useful to consider the conditions under which these paradoxes
could occur in the CPP. The condition $p<0$ can be written 
\beq
\mu>\mu_P(\theta) := 
\frac{r^2 + \ell^2 \cos^2 \theta}{ \ell^2 \sin \theta \cos \theta}.
\label{eq:mumin}
\eeq
Now, in the case of a uniform rod for which $r^2=\frac{1}{3} \ell^2$, we see
that $\mu_p$ is minimised when $\theta=\arctan 2$, 
in which case we can find the
minimum coefficient of friction for which there exists a value for 
$\theta$ for which $p^+<0$, namely $\mumin =4/3$. Thus, for 
$\mu>\mumin$ an interval of angles
$\theta$ exists such that $p^+$ can be negative. 

So what does happen for the falling rod if $\mu>\mumin$? 
This question was analysed in detail
by G\'{e}not and Brogliato \cite{Genot1999} 
(see also \cite{Zhao15} for a modern re-interpretation). 
They show that there is a 
$\mu_c = 8/(3\sqrt{3}) > 4/3 = \mumin$, 
for the case of the uniform rod, 
such that for $\mu<\mu_c$
there can be no initial condition that
approaches $p^+=0$ during slipping, the rod must lift off first. 
But for $\mu>\mu_c$ then there is a thin wedge of initial
conditions $(\theta,\dot{\theta})$ for which a paradox can be reached. 
However they show that there can be no entry during forward slipping 
into regions  with $p^+<0$, the only way to get there is via reaching 
a configuration where simultaneously $b=p^+=0$,
which is
precisely what Or \& Rimon \cite{Or2012} define as 
dynamic jam. As we shall see in Sec.~\ref{sec:5}, the method of
G\'{e}not and Brogliato, involves rescaling time so that
the point $b=p=0$ becomes an equilibrium point in a pseudo phase plain,
which can attract an open set of initial conditions. 
In deference to G\'{e}not, we shall call this special point the
G-spot. What happens after the G-spot is reached is not clear. 

\subsection{Is the paradox due to unrealistic friction models?}
\label{sec:2.2} 

One simple possible resolution of the \pain paradox is that a
coefficient of friction $\mu=4/3$ is unrealistically large, with
typical values for most materials being significantly less than 1
(although values as large as 2 have been proposed for rubber-on-rubber
contacts).  This resolution is specious though because it is easy to
formulate mechanisms for which $\mumin$ is significantly smaller. For
example, if we allow the rod in the CPP to be non-uniform, then
$\mumin$ is a function of the radius of gyration $r$. Taking the limit
that all the mass is concentrated at the centre of mass, then $r \to
0$ and the formula \eq{eq:mumin} shows that $\mumin$ becomes
vanishingly small in this limit.

Another possible resolution is that the Coulomb
friction law is too simplistic and that if a ``more realistic''
friction model is used, the
\pain effect will disappear.  This may be possible for certain cases
and configurations, but seems unlikely in general.  For example, Liu
\textit{et al} \cite{Liu2007} find the paradox to still be present in a
model that has a different dynamic and static coefficient of friction. 
In addition, there is a trivial indeterminacy if, as is typical for
such models, the static 
coefficient of friction is higher than the dynamic one. 

Grigoryan \cite{Grigoryan01} proposes that the CPP can be resolved by
replacing the Coulomb friction law with a smoothed version in which
$\lambda_T$ is a smooth function of $u$ with
large finite slope at $u=0$. Philosophically, this is equivalent
to replacing dry friction with viscous friction for low velocities.
However, he studies only a single rather artificial
model, the so-called \pain-Klein problem (see Sec.~\ref{sec:2.3})
and the resolution doesn't seem to satisfy our definition of
uniform resolvability in the limit that the slope tends to infinity. 
Also, Ivanov \cite{Ivanov03} claims that such friction laws introduce
further non-uniqueness in place of non-existence.   
If instead we replace Coulomb law with a Stribeck law 
(see e.g.~\cite{Rouzic:13}) in
which there is still a jump at $u=0$ but $\mu$ is taken to be a 
smooth function of $u$, then the above calculations for the CPP can
be easily repeated. It is straightforward to show that  
in principle, there is no
impediment to $p^\pm$ becoming negative in this case too. 
The same conclusion holds if $\mu$ is a
function of position $x$ or some internal state variable such as in
rate-and-state friction models (see \cite{Putelat:15} and references
therein).

There is a rich literature on detailed microscopic friction modelling, 
indeed \textit{tribology}\/ is
a field in its own right, which we shall not explore further here. Even
simple models of tangential friction, ignoring the dynamics in
normal directions, can give rise to complex dynamics such as
stick-slip vibrations and ``squeal''; see for example the reviews
\cite{Woodhouse,Kinkaid,Feeny1998}.  Clearly, when dealing with the
\pain paradox, the details of what happens in practice are going to
depend sensitively on the precise friction characteristics.
Nevertheless, it is our contention that the \pain paradox is
fundamentally due to the nature of the coupling between normal and
tangential degrees of freedom, rather than to specific properties
of the friction law.  Therefore, throughout the rest of this paper, for
simplicity, we assume the simple Coulomb law \eqref{eq:Coulomb}, 
with a single coefficient of friction $\mu$.

\begin{figure}
\begin{center}
\includegraphics[width=0.25\textwidth]{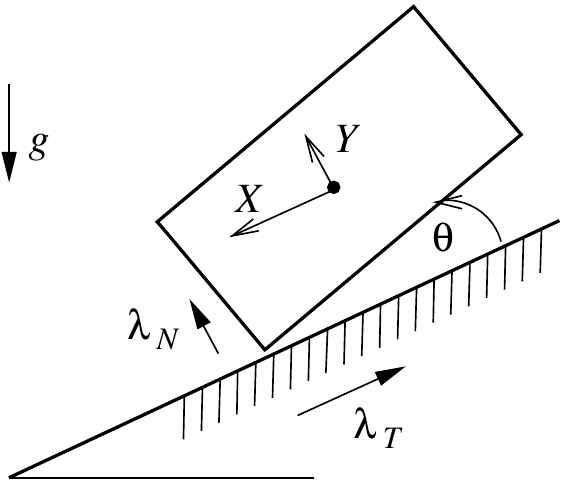}
\end{center}
\caption{The original problem posed by Painlev\'{e} of a slipping block
\label{fig:OPP}}
\end{figure}

\subsection{Other mechanical configurations}
\label{sec:2.3}

In fact, as pointed out by Leine \textit{et al} \cite{Leine2002}, the
original problem studied by Painlev\'{e} in 1895 \cite{Painleve1895}
was not the CPP but that of a planar box slipping down a rough plane
that is inclined at a shallow angle to the horizontal; see
Fig.~\ref{fig:OPP}. The box is assumed to be in point contact at its
lowest corner and to be subject to Coulomb friction.  As Leine \textit{et
  al} show, the simplest model of this problem is actually dynamically
equivalent to the CPP (although see \cite{Mamaev} for a more general analysis)
and the minimum value of friction coefficient required is
identical. In particular, just as with the CPP, if we allow the box to
have inhomogeneous mass distribution, then $\mumin$ can in theory be as
small as we like.

As mentioned in the introduction, the motion of chalk being pushed
across a blackboard is often proposed as a classroom illustration of
the \pain paradox.  In particular, if the chalk is dragged (which
would correspond to an angle $\theta$ in the second quadrant if motion
is to the right in Fig.~\ref{fig:CPP}), then a uniform straight line
is generally seen. However, if the angle is in the first quadrant and
a large force is applied, then the chalk can undergo a series of hops
\cite{Hall}.  Such \textit{sprag-slip oscillations} were reproduced by
Hoffmann and Gaul \cite{Hoffmann} in various models with compliance at
the tip.  These oscillations are at least reminiscent of the \pain
property because lift-off appears to occur despite the chalk being
pressed into the board.  Such an explanation is not sufficient however
because the coefficient of friction for chalk on a blackboard is
likely to be significantly less than $\mumin=4/3$.  However, the
contact mechanics of chalk, a cylinder that is typically only in
contact along part of its end, is likely to be more complex than that
of an ideal rod, and we also have to take account of the body forces
and controls being applied by the teacher. Moreover, we shall also see
in Sec.~\ref{sec:5} that another explanation for the onset of the
hopping motion, namely reverse chatter, can be triggered upon the
transition from stick to slip, even if $p^\pm$ remain positive. Indeed
Nordmark \textit{et al} \cite{paper2} show that an open-loop controller
implementing a vander-Pol-like body force to a simple rod model can
trigger reverse chatter that saturates into limit-cycle motion.

\begin{figure}
\begin{center}
\includegraphics[width=0.7\textwidth]{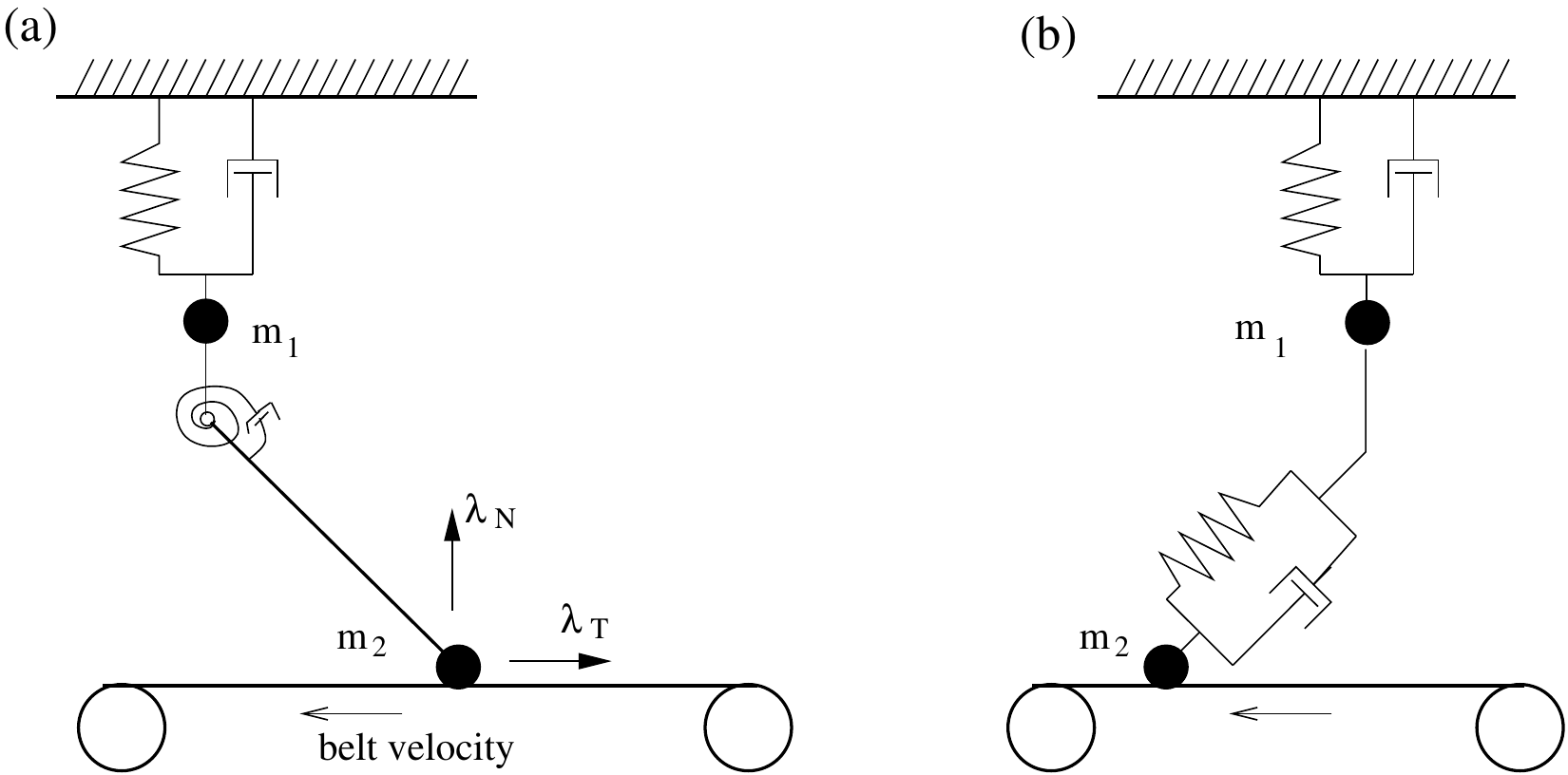}
\end{center}
\caption{The frictional impact oscillator and its simplification, as studied 
in \cite{Leine2002}}
\label{fig:leine}
\end{figure}

A laboratory demonstration of sprag-slip oscillations can be made
through a pin-on-disk rig, if the the pin is allowed to contact
obliquely and to lift off \cite{Ibrahim1994}. Such experimental
apparatus is often used to categorise friction-induced `brake squeal'
vibrations, see for example \cite{Butlin2013}.  Inspired by such
systems, Leine \textit{et al.}~\cite{Leine2002} studied a so-called
frictional impact oscillator, see Fig.~\ref{fig:leine}.
They showed that under suitable choices of parameters the condition
for the \pain property to hold can be written as $\mumin = 2
\sqrt{m_1/m_2}$, which of course can take arbitrarily small values if
the second mass is much heavier than the first. They found that
finite-amplitude stable periodic motion can occur in the \pain
parameter region that comprises phases of free motion, sticking and
slipping. The onset of this motion typically occurs via a subcritical
Hopf bifurcation from the steady sticking solution.

\begin{figure}
\begin{center}
\includegraphics[width=0.7\textwidth]{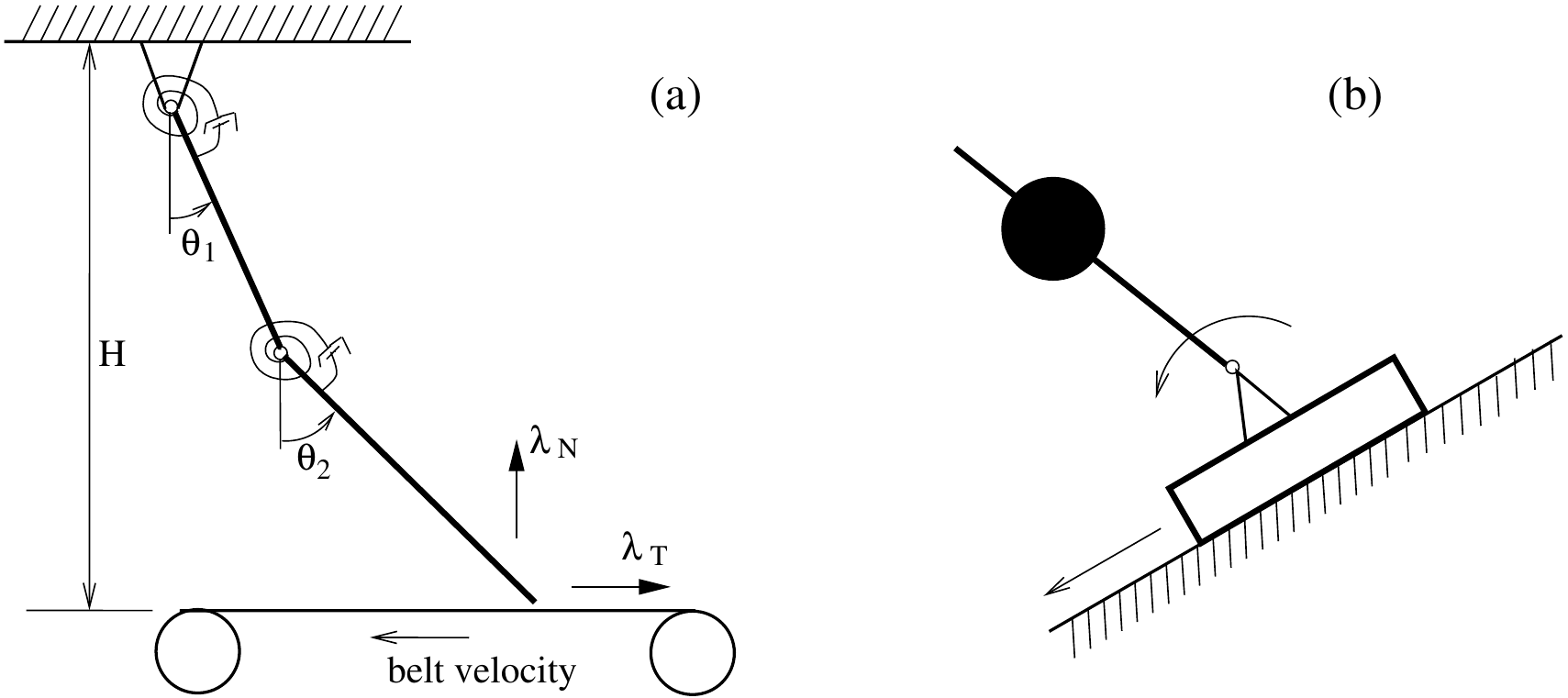}
\caption{(a) The two-link robotic manipulator \cite{Liu2007}. (b)
The inverted pendulum on slider \cite{Or2012}.}
\end{center}
\label{fig:liu_or}
\end{figure}

A related practical manifestation of sprag-slip oscillation is
the hopping motion that is sometimes observed in
finger-like robotic manipulators \cite{Brogliato1999}. Inspired by
this, in a series of papers culminating in \cite{Zhao2008}, Liu, Zhao,
Chen and their co-workers have studied theoretically, numerically and
experimentally the motion of the two-link robotic manipulator depicted
in Fig.~\ref{fig:liu_or}(a). In particular, in \cite{Liu2007} they
found precise formula for $\mumin$, which tends to zero as the height
$H \to 0$.  They were also able to simulate bouncing motion
corresponding to a concatenation of stick, slip, free flight and
impact. That such hopping motion is attributable to the \pain paradox
is now well established in the robotics literature, and work instead
has begun to look at how to control such unwanted behaviour
(e.g. \cite{Elkar:11,Liang}).

There are also possible applications of these ideas in bio-mimicry of
sensing systems. For example, mammals such as rats, have slender
tapered rod-like whiskers that repeatedly sweep and tap on a surface
at oblique angles, see e.g.~\cite{Sullivan2009}. The motion of the tip
of a rat's whisker seems akin to sprag-slip oscillation, but driven
by a regular circular motion from the follicle. It would appear that
the combined effects of stick-slip and lift-off from the tip, fed back
along the whisker enable the rat to sense both texture and compliance
of the surface.

Another experimentally amenable \pain paradox demonstrator was 
proposed by Or \&
Rimon \cite{Or2012}, the so-called
inverted pendulum on a slider, see Fig.~\ref{fig:liu_or}(b).
They were able to find explicit expressions for
initial conditions and parameters 
that would lead to dynamic jam in such a device. 
They found the explicit expression  $\mumin =
\sqrt{(2(1+m_1/m_2)(1+\rho_2/r_2)-1)^2 -1}$, which can be chosen to be
experimentally accessible by choosing appropriate values for the
configuration constants $m_1$, $m_2$ $r_2$ and $\rho_2$.

A less explored, but potentially industrially important area in which the \pain paradox can
apply is in rotating machinery. Here there may be large coupling between normal and tangential
degrees of freedom due to gyroscopic forces.  
For example Wilms \& Cohen
\cite{Wilms1997} report that the \pain effect can be seen the motion
of rotating shaft whose bearing is subject to coulomb friction.
Kozlov \cite{Kozlov} studies a brake shoe problem that exhibits the
\pain paradox, which he claims was resolved by Neimark and co-workers
\cite{Neimark} by introducing longitudinal
and transverse elasticity, see Sec.~\ref{sec:2.4}.

\begin{figure}
\begin{center}
\includegraphics[width=0.35\textwidth]{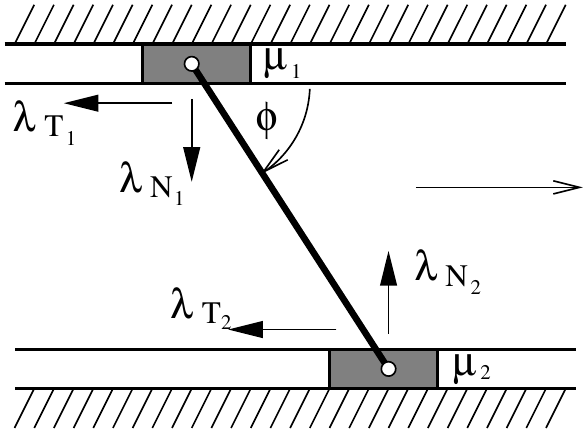}
\end{center}
\caption{The Painlev\'{e}-Klein problem}
\label{fig:Klein}
\end{figure}

\begin{figure}
\begin{center}
\includegraphics[width=0.45\textwidth]{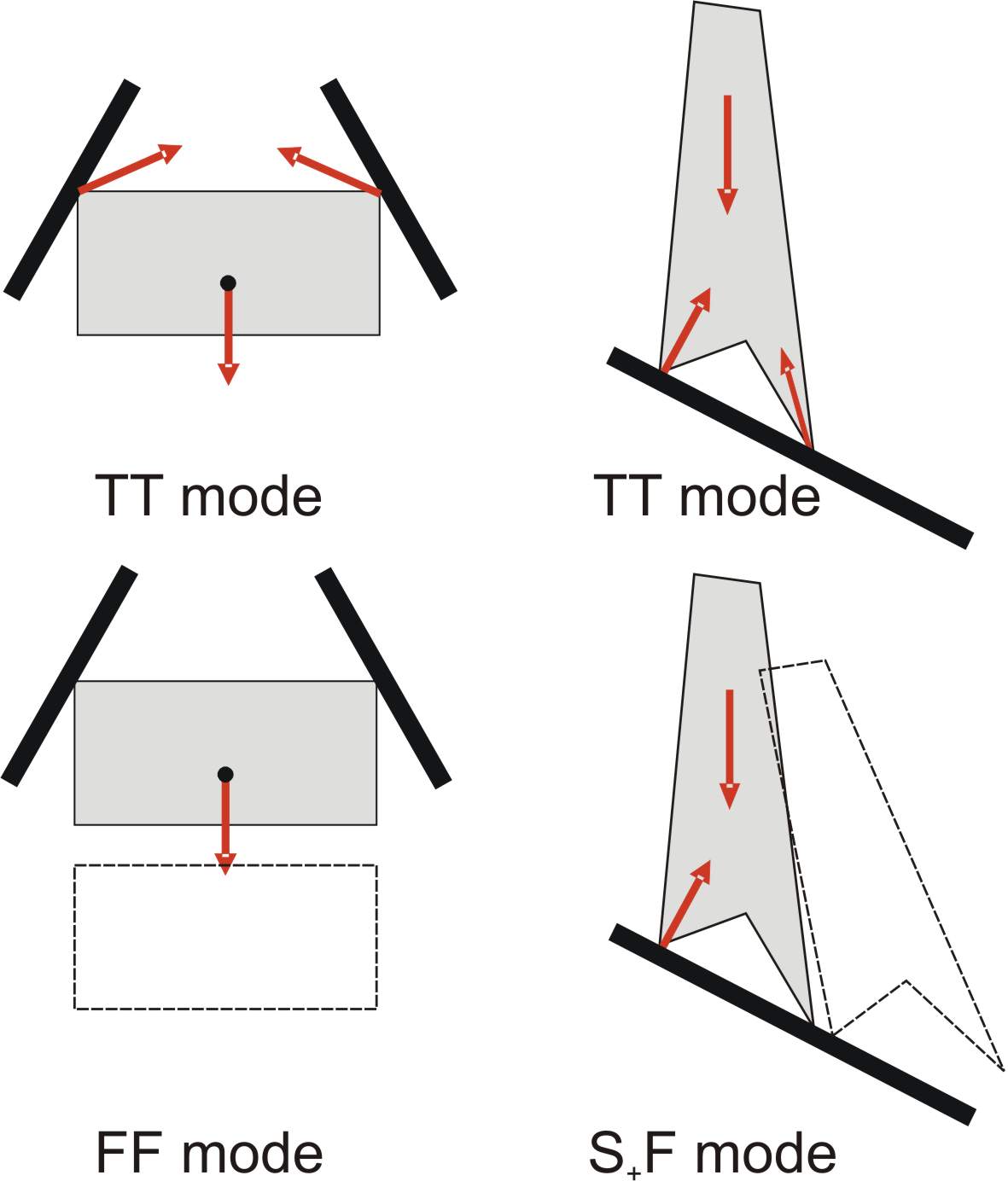}
\caption{(Adapted from \cite{Varkonyi15b}). Two examples of ambiguous
  equilibria involving planar rigid bodies with two contact
  points. (Left) a rigid two-dimensional heavy block between two
  skew walls where both contacts in stick are consistent provided
the coefficient of friction is high enough. (Right) a body resting on
two points in which the left-hand contact is 
slippery, while there is sufficiently high friction at
right-hand contact. Here there is indeterminacy between 
static equilibrium and an accelerating motion in which the left-hand contact
point  slips and the right-hand contact point lifts off. The mode names refer to the notation introduced
in Sec.~\ref{sec:6} below. \label{fig:ambiguity}}
\end{center}
\end{figure}

All of the above examples correspond to cases with a single point
contact. There is less in the literature on the \pain paradox occurring
in problems with multiple contacts.  The Russian literature tends to
use as the canonical model, not the CPP but the so-called
Painlev\'{e}-Klein problem, see \cite{Ivanov03,Neimark,Wagener,Grigoryan01,LeSuan90}. 
A good summary discussion of work on this problem is given in the book by Anh
\cite{LeSuanAn_book}. 
The problem involves a rod that is wedged between two rigid constraints with
Coulomb friction applying at each, see Fig.~\ref{fig:Klein}. In fact,
like the slipping block in Fig.~\ref{fig:OPP}, this problem predates
the CPP, going back to Painlev\'{e}'s original work
\cite{Painleve1895}.  Here it can be shown that there are multiple
cases to analyse depending on whether lift-off is allowed to occur at
 either of the contacts. The simplest case is where both ends are assumed to
always remain in contact, so that the normal forces $\Lambda_{N_{1,2}}$ can take
either sign. Such frictional contacts are termed \textit{bilateral}.
Then, using the notation in Fig.~\ref{fig:Klein}, paradoxes can be shown to 
occur whenever
$$
|\mu_2-\mu_1| \leq \cot \phi.
$$
In particular if the rod is pulled such that a stick-slip
transition is reached, then if $\mu_1=\mu_2$ one can't decide
which end will slip first. Ivanov \cite{Ivanov03} analyses
cases where either or both of the constraints are 
unilateral, showing that there is a much richer complexity of
possibilities, depending on the relative sizes of the two
coefficients of friction. In general, we shall consider only
unilateral constraints in what follows and, for the most part, mechanisms
with just a single contact point. Extension to multiple contact points 
is the subject of Sec.~\ref{sec:6}, where we will specifically
point out several different cases of nonuniqueness,
as illustrated in Fig.~\ref{fig:ambiguity}.

\begin{figure}
\begin{center}
\includegraphics[width=0.75\textwidth]{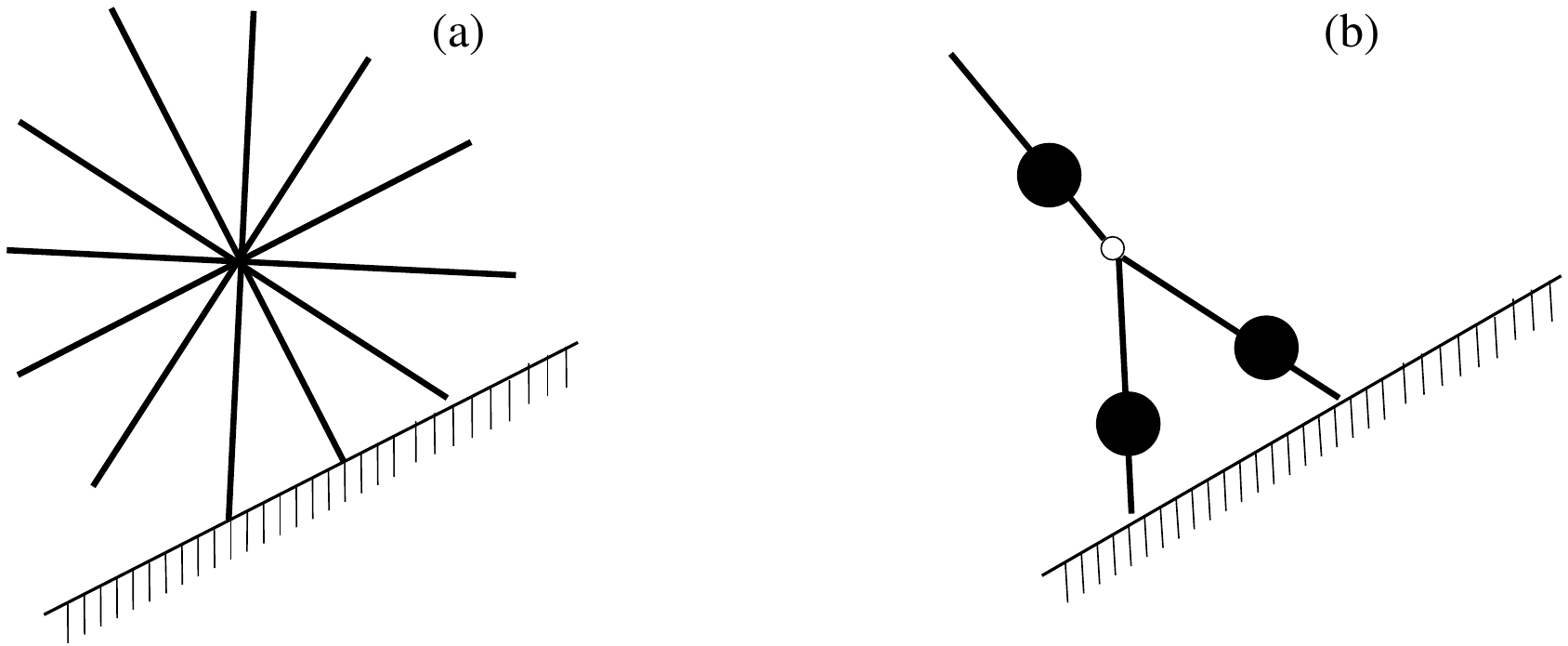}
\end{center}
\caption{(a) The rimless wheel and (b) the compass biped configurations studied
in \cite{Or2014,GamosOr}.} 
\label{fig:compass}
\end{figure}

Or \cite{Or2014} studied practical configurations 
with multiple points that can exhibit
the \pain paradox, simple passive walking mechanisms such as the rimless wheel
and the compass biped, see Fig.~\ref{fig:compass}. 
He argues that many walking models assume frictional sticking
contact of the leg with the ground. Allowing perturbations
that involve foot slippage, he shows that 
regular gait periodic solutions can be subject to an instability
that is closely related to dynamic jam. Dynamical consequences of
this instability and the ensuing stable dynamical behaviours are considered
in detail in \cite{GamosOr}. 

\subsection{Resolutions of the paradox; regularisation and 
impact mechanics} 
\label{sec:2.4}

Various attempts to resolve the \pain paradox have taken place in the
intervening 120 years since Painlev\'{e} first published his
work. These attempts took on new vigour in the 1990s due to the
development of rigorous methods using the complementarity framework
for rigid body mechanics and the theory of differential inclusions; see
for example \cite{Stewart2000} for a review. One idea, due 
to the  influential French
mathematician Moreau \cite{Moreau1988}, is to solve the problem via
simulation, using specific time-stepping numerical
discretisation schemes designed for linear complementarity problems.
Using this idea Stewart \cite{Stewart1997,Stewart1998} provided a resolution
in the form of a rigorous proof that a time-stepping schemes exist that are
well posed and, by taking the zero-stepsize limit, one has a proof
that mechanics in the \pain regime is \textit{consistent}. That is, there
is always a well defined forward-time evolution of the dynamics from
any reachable configuration.  However, this approach does not deal
with the problem of \textit{indeterminacy} that is, one can have
non-uniqueness in the forward time dynamics. 
Stewart's result also does not resolve the paradox in the sense of the three 
Questions posed in the introduction
as it does not consider the nature of the solution in the 
limit as the stepsize tends to zero. 

Another way to resolve the Painlev\'{e} paradox is to break the
formalism of rigid body mechanics and to introduce some new physics,
for example by including dynamics in the normal direction, thus
smoothing the problem. This approach seems to be prominent in the
Soviet literature, following the publication of the Russian
translation of Painlev\'{e}'s work in 1954 (see
e.g.~\cite{LeSuan90,Ivanov03,Neimark,LeSuanAn_book} and references
therein).  For example, Neimark and Smirnova \cite{Neimark} propose
that the normal force should be considered as a dynamical variable
that evolves on a fast timescale. They argue that the dynamics of the
Painlev\'{e}-Klein problem would then feature ``contrast structures''
by which they mean two-timescale dynamics that can converge to
periodic solutions with well-defined rapid jumps in tangential
velocity. Without this regularisation they claim that only constant
acceleration solutions can be found in the Painlev\'{e}-Klein problem.
This general approach can be referred to as \textit{contact
  regularisation}, which treats the rigid body limit of a compliant
formulation as a singular perturbation problem. A good general
discussion of contact regularisation and its application to the
Painlev\'{e}-Klein problem in particular can be found in
\cite{LeSuanAn_book}.  In Sec.~\ref{sec:4} below we show how such
contact regularisation in the normal direction 
can lead to answers to Questions 2 and 3 by
taking the limit that the stiffness, following arguments in
\cite{paper2}.

Most authors agree that a complete understanding of the Painlev\'{e}
paradox requires consideration of impact. That is, what would happen
in the CPP if the rod is first dropped onto the surface so that it
first makes contact with $v<0$ and $b>0$. In general, contact will not
be maintained, but the body will bounce.  Even before the advent of
classical mechanics, a great deal of effort was made to understand
collisional impact, see \cite[Appendix A]{Stronge2000} for an
historical account.  Within a rigid-body framework, the loss of energy
in the impact event is usually modelled as a zero-time process
involving impulsive forces.  Impacts for which there is no coupling
between tangential and normal degrees of freedom during contact are
most simply captured by a Newtonian restitution law \beq v^+ = - r
v^-, \qquad r\in [0,1].
\label{eq:Newtonian}
\eeq
A completely elastic (conservative) collision corresponds to 
a \textit{coefficient of restitution} $r=1$, and a completely inelastic
case to $r=0$. Of course, even this classical law is an approximation, as
Newtonian coefficients of restitution are approximations that depend 
not only on the properties of the contacting materials but
on how the geometry of the impacting bodies
allow wave energy to be dissipated, see e.g.~\cite{Melcher,Szalai2}.

The \pain paradox involves {\em oblique} impacts, which
involve instantaneous changes in the tangential as well as
normal velocity. Unlike purely normal impacts, it appears 
from the literature that 
different modelling choices can be made in order to
resolve oblique impacts. 
within a rigid-body framework.  
First, it is tempting to simply
ignore the tangential dynamics during impact and apply
\eq{eq:Newtonian} in the normal direction.  However, this can lead to
an increase in energy during impact
\cite{Kane1984,Chatterjee1998}. Another obvious idea is to introduce a
second model parameter, --- a ``transverse coefficient of
restitution'' \cite{Pfeiffer2000} or a fixed ``impulse ratio'' between
the tangential and normal velocity jumps \cite{Brach1991} but this can
similarly be shown to lead to energy gain in Painlev\'{e} paradox
situations \cite{Burns}. Chatterjee and Ruina \cite{Chatterjee1998}
discuss which closed form expressions for impact lead to 
oblique impacts that are energetically consistent.

As pointed out in \cite{Chatterjee1998}, the problem of many
simplistic approaches to modelling oblique impact is that they ignore the fact
that during the impact process a transition from slip to stick can
occur. In fact, as argued by Stronge \cite{Stronge2000}, see also
Sec.~\ref{sec:3.2} below, consistent impact laws can only be reached
in all circumstances by solving the dynamic problem of compression
followed by restitution, in a rapid timescale in which the impulsive
forces become of $O(1)$, fully resolving any transitions between
slip. This is a form of contact regularisation that makes impact
resolvable by passing to the limit that its duration is
infinitesimal.  The question then arises as to what is the analogue of
the coefficient of restitution for such a process.  Or, more
precisely, when do we decide that the restitution phase has
terminated?  Various alternatives are possible, with the simplest one,
based on a condition on normal velocities akin to \eq{eq:Newtonian}
being easily shown to be inconsistent, see \cite{Sanders2013} for a
comparison.  One possible resolution is to use Poisson's kinetic
restitution law \cite{Poisson1811}, see also
\cite{Keller1986,Batlle1993} that supposes there is a ratio between
the normal impulse in compression to that in restitution.
Alternatively, Stronge \cite{Stronge1990} proposed an \textit{energetic}
restitution law that considers the ratio of normal work done in
compression to that in restitution. Stronge's law has the benefit that
by construction, it is energetically consistent that is, the impact
must be dissipative if the energetic coefficient of restitution is
strictly less than unity, and is necessarily conservative if it equals 
unity. Nevertheless, energetic consistency can also be proved to hold for the
Poisson law \cite{Ivanov93,Chatterjee1998}.  When the \pain property
holds, both laws give the possibility of impact without collision (IWC)
\cite{Genot1999,Stewart2000}. That is, where a finite outgoing normal
velocity ensues from a zero incoming velocity.  As we shall see in
Sec.~\ref{sec:3.2}, such events can provide a resolution to the
inconsistent case of the Painlev\'{e} paradox.

There are relatively few studies that analyse or simulate 
configurations exhibiting the \pain property with models that incorporate
both continuous contact and impact.  
For example, in one of the first papers to analyse the \pain phenomenon from
the perspective of non-smooth bifurcations, Leine 
\textit{et al.}~\cite{Leine2002} assumed a 
coefficient of restitution equal to zero. This was improved
upon by Liu \textit{et al.} \cite{Liu2007} who 
simulated similar hopping motion using a non-zero Poisson coefficient 
of restitution $r_p$, but their analysis of periodic motion is 
restricted to the case case $r_p=0$.

In contrast, Stronge and co-workers \cite{ShenStronge,Stronge13,Stronge15} 
consider in detail the transitions
that occur during impact, but ignoring the case of IWC.
They find that there are no paradoxes. Indeed, this should
not be surprising, because, 
as shown in \cite{paper1,Stronge1990,Djerassi2009} 
explicit expressions for the outgoing velocity 
in terms of the incoming velocity can be obtained for each of the Newtonian,
Poisson and Stronge restitution laws, for all possible itineraries of slip
and stick during compression and restitution.   

\section{General formulation for planar single contact case} 
\label{sec:3}
%
%
%
This section is devoted to the answering Question 1 in the restricted
case of planar systems with a single point of contact, without introducing
any additional ingredients to the rigid-body formulation. We shall deal 
with typical cases, given by open (generic) conditions on parameters.
Cases that are on the boundary of these open regions 
are dealt with in Sec.~\ref{sec:4}.

Consider a planar mechanical system  with a single point whose dynamics
is governed by the Lagrangian system
\beq
M\left(q ,t\right)\ddot{q}  =f(q,\dot{q},t)+
\lambda_T  c(q,t)+\lambda _{N} d(q,t).  
\label{eq:lagrange}
\eeq
Here $q \in \Rset^n$ is a vector of generalised co-ordinates, with
$\dot{q}$ the vector of corresponding generalised velocities, $f$ 
contains all body forces, including potentially both conservative and
dissipative forces, and $\lambda_T$ and $\lambda_N \geq 0$ 
represent the magnitudes
of tangential and normal forces respectively that act in the directions
corresponding to the generalised co-ordinate vectors $c(q)$ and $d(q)$.
$M(q)$ is a mass matrix which we assume to be positive definite for
all admissible configurations $q$. Each
of $M$, $f$, $c$ and $d$ is assumed to be a 
sufficiently smooth function of its arguments.   

Let $(x(q),y(q))$ be the co-ordinates associated with the $c$ and $d$ 
directions so that the constraint is given by $y\geq 0$, and let
$\dot{x}=u$ and $\dot{y}=v$ be the tangential and 
normal velocities respectively. Then, we can project \eqref{eq:lagrange}
onto tangential and normal directions to obtain the scalar equations  
\begin{align}
\dot{u}& =a\left( q,\dot{q},t\right) +\lambda_{T}A\left( q,t\right)
+\lambda_{N}B\left( q,t\right) ,  \label{eq:udynamics} \\
\dot{v}& =b\left( q,\dot{q},t\right) +\lambda_{T}B\left( q,t\right)
+\lambda_{N}C\left( q,t\right) ,  \label{eq:vdynamics}
\end{align}
where the scalars $a$, $b$, $A$, $B$, $C$, $D$ are given by 
\begin{align*}
& a = c^T f, \quad b = d^T f, \quad 
A=c^T M^{-1} c,\\
& B=c^T M^{-1} d
\quad \mbox{and} \quad C=d^T M^{-1} d.
\end{align*}

The scalars $\lambda_N$ and $\lambda_T$ are Lagrange multipliers that must
be solved for under different assumptions on the mode of motion. 
Whenever
$y>0$ or $v>y=0$, we have  free motion for which necessarily $\lambda_N=\lambda_T=0$. 
During contact $y=v=0$, we suppose that 
Coulomb friction \eqref{eq:Coulomb} applies. 

Note that because $AC-B^2$ is the determinant of a 
$2\times 2$ submatrix of $M^{-1}$ and $A$ and $C$ are diagonal
elements in an appropriate basis, then the positive definiteness of
$M$ implies that 
\beq
A>0, \quad C>0, \quad AC-B^2>0, 
\label{eq:ABCD_inequals}
\eeq
whereas $a$, $b$ and $B$ are in general not sign constrained.
Moreover from \eqref{eq:ABCD_inequals}, simple algebraic manipulation shows
that  at most one of 
\beq
\mu A -B, \quad \mu A + B, \quad C- \mu B, \quad C + \mu B 
\label{eq:nonpositive}
\eeq
can be non-positive, which will be important in what follows. 
The case $B=0$ corresponds to there being no coupling
between the normal and tangential forces during contact. The \pain paradox
can occur whenever $|B|$ is sufficiently large. Note, from the CPP 
example \eqref{eq:CPP}, in the case of a uniform rod, nondimensionalisation
using length scale $\ell$, time scale $\sqrt{\ell/ g}$ and mass scale
$m$, gives
$a= -\dot{\theta^2} \cos \theta$, $b= \dot{\theta^2} \sin \theta -1$,    
$A = 1 + 3 \sin^2\theta$, 
$B = 3 \sin \theta \cos \theta $, 
$C = 1 + 3 \cos^2 \theta$.

\subsection{Consistency of contact motion} \label{sec:3.1}

Sustained contact occurs when $y=v=0$, from which we can distinguish 
four generic possible modes of motion; lift-off into {\bf f}ree motion ($F$), 
s{\bf t}ick ($T$) and {\bf s}lip either in the positive $u$ ($S_+$), or negative $u$ 
($S_-$) direction. To determine which mode is consistent for 
any configuration, we can apply a 
three step algorithm, see Table \ref{tab:consistency}: 
\begin{enumerate}
\item check kinematic admissibility; 
\item find contact forces and accelerations using equations of motion and 
equality constraints; 
\item check consistency conditions. 
\end{enumerate}

Let us now consider the details of the consistency analysis 
for the four contact modes:
\paragraph{Free motion}
occurs when $\lambda_N=\lambda_T=0$, which implies $\dot v = b$. The consistency condition $\dot v=0$ is satisfied if $b>0$.
\paragraph{Positive slip} occurs when 
$y=0$, $v=0$, $\lambda_N>0$ and $u \geq 0$. We now have 
the full friction force so that $\lambda_T=-\mu\lambda_N$ and hence to
sustain contact we must have
\begin{equation}
    \dot{v} = b+ (C-\mu B)\lambda_N =0. \label{eq:normaldynamics-slip+}
\end{equation}
We can define the positive \pain parameter 
\beq
p^+ := C-\mu B,
\label{eq:pain_par+}
\eeq
and rewrite \eqref{eq:normaldynamics-slip+} as
\begin{equation}
    \dot{v} = b+ p^+\lambda_N =0, \label{eq:normaldyn_02_slip+},
\end{equation}
which implies
\begin{equation}
  \lambda_N=-\frac{b}{p^+}, \qquad   
    \dot{u} =a -b \frac{(B-\mu A)}{p^+}. \label{eq:udotslip+}
\end{equation}
Thus, the consistency condition $\lambda_N \geq 0$ becomes that 
$b$ and $p^+$ should have opposite sign 
(with $p^+ =0$ leading to $\lambda_N$ being undefined). We 
say that the \pain paradox for positive slip occurs when $p^+<0$. 
Note that the additional consistency condition for the case $u=0$ is
that $\dot{u}$ given by the second equation in \eq{eq:udotslip+} should
be positive, that is
\begin{equation}
    a -b \frac{(B-\mu A)}{p^+}>0. \label{eq:udotslip+>0}
\end{equation}

\paragraph{Negative slip} similarly occurs when 
$y=0$, $v=0$, $\lambda_N>0$ and $u \leq 0$, $\lambda_T=\mu\lambda_N$ and 
\begin{equation}
    \dot{v} = b+ p^{-}\lambda_N =0. \label{eq:normaldynamics-slip-}
\end{equation}
where the \pain parameter for negative slip is
\beq
p^- := C+\mu B.
\label{eq:pain_par-}
\eeq
So, we have 
\begin{equation}
  \lambda_N=-\frac{b}{p^-}, \qquad   
    \dot{u} =a -b \frac{(B+\mu A)}{p^-}, \label{eq:udotslip-}
\end{equation}
and hence the consistency condition for negative slip is that $b$ and
$p^-$ should have opposite sign, and we identify the case $p^-<0$ as
representing the \pain paradox for negative slip. Again there is an extra
consistency condition on $\dot{u}$ in the case $u=0$, that is, 
\begin{equation}
a -b \frac{(B+\mu A)}{p^-} <0. \label{eq:udotslip-<0}
\end{equation}

\begin{figure}
    \centering
   \includegraphics{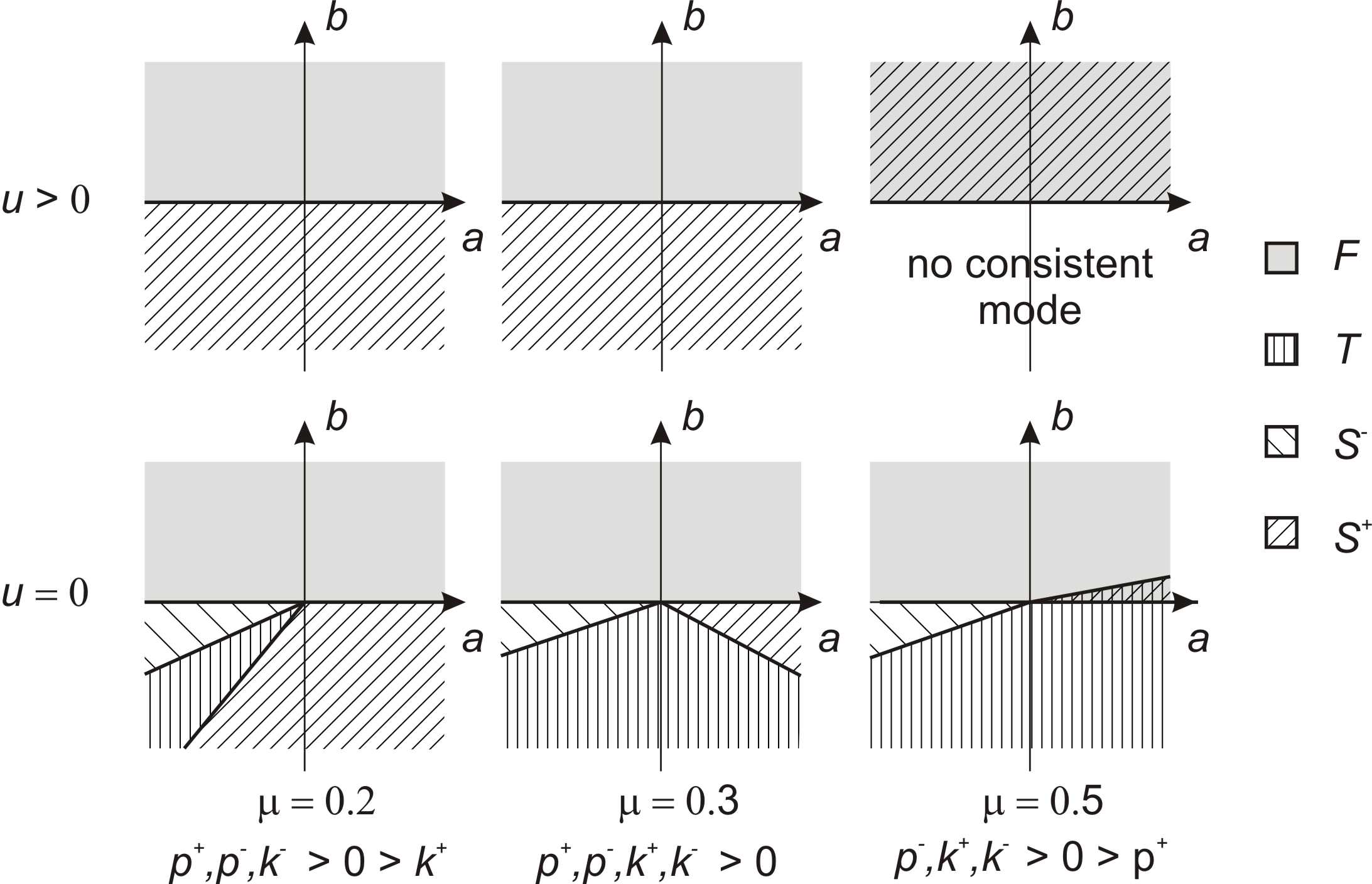}
    \caption{
Consistency regions of the four contact modes in the plane of 
parameters $a$ and $b$ at three values of the coefficient of friction 
$\mu$, for sustained contact $y=v=0$ with positive tangential velocity ($u>0$, top) and with stationary contact ($u=0$, bottom). 
The case  illustrated is for
$A=1.1$; $B=0.3$; $C=0.1$. The largest value of $\mu$ induces $p^+<0$ for
which there are regions of 
indeterminacy and inconsistency in the parameter plane.}
    \label{fig:consistency}
\end{figure}

\paragraph{Stick} represents a mode for which 
$y=0$, $v=0$, $\lambda_N>0$, $u=0$, and 
$|\lambda_T|<\mu\lambda_N$. In order to sustain stick we must have
$\dot{u}=\dot{v}=0$ from which we can explicitly obtain 
\begin{equation}
\left( \lambda _{T},\lambda _{N}\right) =\left( \frac{bB-aC}{AC-B^{2}},\frac{
aB-Ab}{AC-B^{2}}\right).  \label{eq:stickforces}
\end{equation}
The consistency condition $|\lambda_T| < \mu \lambda_N$ to remain in stick, can
thus be written as
\beq
    a(C-\mu B)+b(\mu A-B)<0 , \quad -a(C+\mu B)+b(\mu A+B)<0. 
\label{eq:frictioncone}
\eeq
By analogy with the 3D case, the conditions 
\label{eq:frictioncone}
are sometimes said to
define the interior of the \textit{friction cone} in the
$(a,b)$-plane.
This region is represented by the vertically hashed area 
in Fig.~\ref{fig:consistency}. Note the shape of the cone
depends on the signs of $p^\pm$ and the additional parameters 
\beq
k^+ := \mu A- B, \qquad k^- := \mu A+ B.
\label{eq:pain_pars2}
\eeq
(Note a change in sign convention for the definition of $k^+$ from \cite{paper1}
where it was called $-k_T^+$.)

\begin{table}
\begin{center}
\begin{tabular}{|p{2.5cm}|p{3.75cm}|p{3.75cm}|p{3.0cm}|p{3.0cm}|}
\hline
Mode & pos.~{\bf S}lip ($S_+$) & neg.~{\bf S}lip ($S_-$) & s{\bf T}ick ($T$) 
&  lift of{\bf F} ($F$) \\
\hline
kinematic admissibility & $y=v = 0$ and $u \geq 0$ & $y=v = 0$ and $u \leq 0$ & $y=v = 0$ and $u = 0$ & $y>0$; or $y=0$ and $v \geq 0$ \\
\hline
equality constraints & $\lambda_T=-\mu\lambda_N$ and $\dot v=0$  &  $\lambda_T=\mu\lambda_N$ and $\dot v=0$ & $\dot u =\dot v=0$ & $\lambda_N = \lambda_T = 0$ \\
\hline
consistency &$\lambda_N\geq0$; and if $u=0$: $\dot u>0$ & $\lambda_N\geq0$; and  if $u=0$: $\dot u<0$  &$\lambda_N\geq0$; and $|\lambda_T|\leq\mu \lambda_N $  & if $y=v=0$: $\dot v >0$  \\	
\hline
\end{tabular}
\caption{Constraints of the four contact modes.}
\label{tab:consistency}
\end{center}
\end{table}

\bigskip

All four parameters $p^\pm$, $k^\pm$ are positive in 
the case $B=0$ and, owing to \eqref{eq:nonpositive}, at most one of them 
can be negative in generally. 
If one the $k$ parameters is negative then
we have the case illustrated in the bottom left panel of
Fig.~\ref{fig:consistency} in which the conditions for stick are
completely contained within one quadrant.  If instead one of the $p$
parameters is negative then we have the case in the bottom right panel
in which the stick region extends into the lift-off zone $b>0$.
In this latter case there is 
a range of $a$-values for which stick and slip are consistent even if $b>0$.

\begin{table}
\begin{center}
\begin{tabulary}{\textwidth}{|R|J|J|J|}
\hline
   &I: \: $u>0$ & II: \: $u=0$ & III: \:  $u<0$ \\
   \hline
   (i): $b>0$, $p^+>0$, $p^->0$  &$F$&$F$&$F$\\
    (ii): $b>0$, $p^+>0$, $p^-<0$  & $F$ & $F$  or  ($F$ and $T$ and $S_-$) 
& $F$ and $S_-$ \\
     (iii): $b>0$, $p^+<0$, $p^->0$  & $F$ and $S_+$ & 
$F$ or ($F$ and $T$ and $S_+$) & $F$ \\
     \hline
  (iv): $b<0$, $p^+>0$, $p^->0$    & $S_+$ & $T$ or $S_+$ or $S_-$ & $S_-$ \\
    (v): $b<0$, $p^+>0$, $p^-<0$  & $S_+$ & $S_+$ or $T$ &---\\
     (vi): $b<0$, $p^+<0$, $p^->0$ &--- & $S_-$ or $T$ & $S_-$ \\
     \hline
\end{tabulary}
\caption{Consistency of the contact modes depending on the signs of 
$b$, $u$, $p^+$, $p^-$. Here `or' is exclusive and which mode occurs depends on other inequality conditions
between $a$, $b$, $A$, $B$ and $C$; specifically, stick occurs if and only
$a$ and $b$ lie in the interior of the friction cone
given by \eq{eq:frictioncone}.  
 However `and' is inclusive and represents indeterminacy. 
A `---' symbol is used to denote
cases where there is no consistent mode.} 
\label{tab:summary}
\end{center}
\end{table}

The results of this consistency analysis are 
summarised in Table \ref{tab:summary} and illustrated in Fig.~\ref{fig:consistency}. 
Note the 
{\bf inconsistency} in cells I(vi) and III(v) of the table. 
These cases
are symmetric with respect to each other under the transformation 
$x \to -x$, $B \to - B$, which maps forward to backward slip, 
so without loss of generality we consider the case for positive slip; 
$y=v=0$, $b<0$ $p^+>0$, $u>0$. Here, stick is not possible since
$u>0$, nor is lift-off into free motion, because the free acceleration
$b$ is in the direction towards the constraint surface $y=0$. Negative
slip is not possible because $u>0$, and finally positive slip is not possible
because $\lambda_N$ given by \eqref{eq:slip_normal_force} would be negative.
To resolve what must happen in such cases, in practice we need to consider
the possibility of an IWC, which will do in the next subsection.

There are also two different types of
{\bf indeterminacy}, comprising four cases in all. The first type is
indeterminacy between slip and lift-off in cells I(iii) and III(ii),
with these two cases being symmetric with respect to each other under
the transformation that maps positive slip to negative slip.  So,
without loss of generality, consider positive
slip; that is $y=v=0$, $b>0$, $p^+>0$, $u>0$. Here, stick is not
possible because $u>0$ but lift off is possible, as is positive slip,
with $\lambda_N$ given by \eqref{eq:slip_normal_force} being positive
because it is the ratio of two negative quantities.  The other type of
indeterminacy is between liftoff, stick and slip, in cells II(ii) and
II(iii) where $b>0$ and $a$ is such that we lie inside the friction cone
(that is, inside the multi-shaded region of the  bottom-right panel of 
Fig.~\ref{fig:consistency}). Here lift-off is possible because $b>0$. 
Additionally though, a positive normal force $\lambda_N$ can be found such
that the conditions for stick are satisfied. In addition, a larger normal force
$\lambda_N$ exists such that slip is possible for the direction corresponding
to the parameter $p^\pm$ that is negative.  
These cases of indeterminacy cannot be
resolved in general (in the sense of Question 1 in the introduction),
but can be resolved in terms of their stability (in the sense of
Question 2) see Sec.~\ref{sec:4} below.

\subsection{Impact}
\label{sec:3.2}

As discussed in Sec.~\ref{sec:2.4}, the discussion of the \pain paradox
in a rigid body framework requires the inclusion of an impact law. 
Indeed,
if contact regularisation is included in the normal direction, the necessity of
IWC is easily concluded (see e.g.~\cite{Neimark}).
Following \cite{Stronge2000,paper1}
we shall introduce a formalism for the impact phase as something
occurring on an asymptotically faster timescale $\tau=t/\varepsilon$ in
which 
contact forces become asymptotically large $(|\lambda_T|,\lambda_N=\mathcal{O}(\epsilon^{-1}))$ 
such that velocities $(u,v)$
(and more generally $\dot{q}$) can vary by an $O(1)$ amount, but the
generalised coordinates $q$ remain constant. We shall introduce a
notation that the pre- and post-impact velocities are represented using
superscripts $^-$ and $^+$ respectively.

\begin{figure}
\begin{center}
\includegraphics[width=0.7\textwidth]{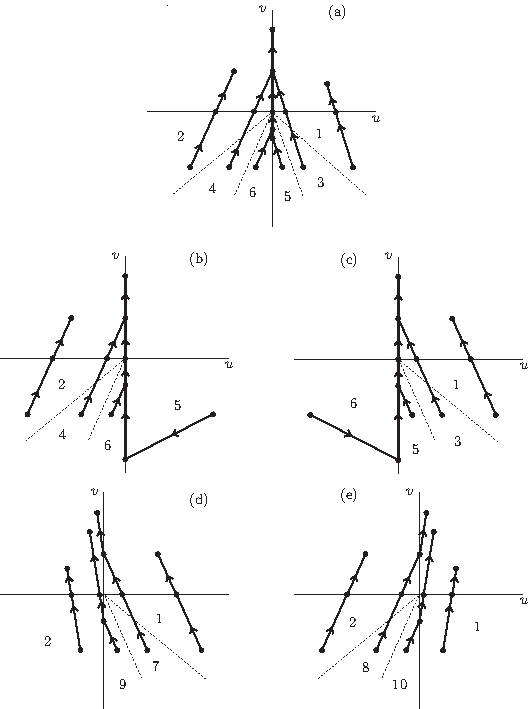}
\end{center}
\caption{After \cite{paper1}, reproduced with permission.
Graphical description of the impact process, in:
the typical  case (a) $p^+,p^-,k^+,k^->0$; the
\pain cases (b) $p^+<0$, (b) $p^-<0$; or the slip 
reversal cases (c) $k^-<0$, (d) $k^+<0$. 
In each case the mapping is depicted from an incoming velocity pair
$(u^-,v^-)$ with $v^- \leq 0$ to an outgoing pair $(u^+,v^+)$ with 
$v^+>0$.  The dashed lines represent boundaries between 
regions in $(u^-,v^-)$-space,  represented by a different numerical symbols,  
in which the map takes a different functional form.}
\label{fig:impactmap}
\end{figure}

Then, if we define the rescaled contact forces $\Lambda_{T,N}= \varepsilon \lambda_{T,N}$ 
then 
from \eqref{eq:udynamics}, \eqref{eq:vdynamics} we get in
\begin{align}
\frac{du}{d\tau} & = \Lambda_{T}A +\Lambda_{N}B +O(\varepsilon), \label{eq:uimpulse} \\
\frac{du}{d\tau} & = \Lambda_{T}B +\Lambda_{N}C + O(\varepsilon), \label{eq:vimpulse}
\end{align}
where $A$, $B$ and $C$ are now constant during the impact process. 
Moreover, we can replace time $\tau$ with units of the \textit{normal impulse} \cite{Stronge2000}
$$
I_N = \int \Lambda_N d \tau 
$$ 
which is a monotonically increasing function of $\tau$. Then we get
\begin{align*}
u^\prime & = (\Lambda_{T}/\Lambda_N) A + B +O(\varepsilon), \\
v^\prime & = (\Lambda_{T}/\Lambda_N)B + C + O(\varepsilon),
\end{align*}
where a prime represents $\frac{d}{d I_N}$.
These equations, after substitution from the 
Coulomb friction law \eqref{eq:Coulomb}
with $\lambda_{T,N}$ replaced by $\Lambda_{T,N}$, lead to leading order to 
explicit affine equations for $u(I_N)$ and $v(I_N)$. 
In fact,  as shown in \cite{paper1},
the motion is always along straight lines in the $(u,v)$-plane, 
with corners occurring at transitions between slip and stick during the impact process; 
see Fig.~\ref{fig:impactmap} for examples.

The impact process can then be defined as a composite mapping 
\begin{equation}
(u^-,v^-) \mapsto_{\mbox{compression phase}} (u^*,0) 
\mapsto_{\mbox{restitution phase}} (u^+,v^+), 
\label{eq:impactmap}
\end{equation}
where in each of the compression and restitution phases one needs to account
for possible transitions from slip to stick.
It is possible to then define 
a composite closed from expression for the impact map under different 
combinations of the signs of $p^+$, $p^-$, $k^+$ and $k^-$ defined in
\eqref{eq:pain_par+}, \eqref{eq:pain_par-} and \eqref{eq:pain_pars2}, 
and different 
conditions on the ratio $u^-/v^-$ between initial velocities. 
The 
results are summarised in Fig.~\ref{fig:impactmap}; detailed
calculations under the assumption of an energetic impact law
in \cite{paper1}. In that case we deem an impact to be over when
the normal kinetic energy gained during restitution is
$-r^2$ times the normal kinetic energy lost during compression.
This defines an energetic coefficient of restitution $r=r_e$.
Similar calculations can be carried out explicitly
for the Poisson impact law (see e.g.~\cite{Chatterjee1998}), the only
difference being the point at which the restitution phase is deemed to
be over. In that case, restitution is deemed to end when the 
normal impulse in $I_N = \int \Lambda_N$ gained during restitution is $r$ times
the normal impulse gained during compression, which 
defines a Poisson coefficient of restitution $r=r_p$

Note that the assumption that the \textit{same} Coulomb friction law
should apply during the impact phase as in the $O(1)$-timescale motion
is a modelling assumption that doesn't necessarily follow. 
For example, when modelling
impact in a so-called superball, it has been suggested that a friction
law is used where the transition between stick and slip occurs not at $u=0$ 
but for some non-zero $u>0$~\cite{Cross15}.
In truth, 
frictional forces are temperature, timescale and spacescale dependent, 
see e.g.~\cite{Woodhouse}.  The impact process \eqref{eq:impactmap} 
can in principle be defined under different friction laws, or 
indeed for a different value of
the coefficient of friction $\mu_I$ than for the $O(1)$-timescale
motion. 

Note that in general $v^-<0$, but that in parameter regions 5(b) or 
6(c) in the figure, the impact equations also have a solution 
if $v^-=0$. These regions correspond precisely to when either of the \pain
parameters $p^+$ or $p^-$ is negative and the pre-impacting motion is
in slip of the correct sign ($u>0$ or $u<0$ respectively).  In
particular these two parameter regions are precisely where an impact
can occur with $v^-=0$, that is an \textit{impact without collision}.
Note these cases map
to the inconsistent and the indeterminate cases of 
Table \ref{tab:summary} with $u\neq 0$. 
Thus, we have at least one forward solution in
all cases.

Also note from the Fig.~\ref{fig:impactmap} 
that impact can result in the phenomenon of
\textit{slip reversal}, that is where $u^+u^-<0$ so that the body will 
enter impact with slipping in one direction and exit slipping
in the other. This phenomenon occurs in parameter regions 7-10 of
the figure which occurs when one of the 
parameters $k^+$ or $k^-$ is negative.

\subsection{Chattering and inverse chattering} 
\label{sec:3.3}

Chattering, also known as the Zeno phenomenon, is the process by which
an infinite sequence of impacts occurs in a finite period of time. In
single degree-of-freedom mechanical systems, such a sequence can
easily occur whenever there is a coefficient of restitution less than
unity and acceleration that is towards the contact for a sufficiently
long period of time. The canonical example of such an impact sequence
occurs when dropping an elastic ball on a rigid floor, and can also
occur in general impact oscillators, see
\cite{Springerbook,PetriNordmark} and references therein. For systems 
with oblique impacts, chattering sequences can 
also converge in reverse time.

To understand this phenomenon, 
following \cite{paper2} we define a \textit{chattering sequence} as 
a rapid sequence of impacts interspersed with brief intervals of
free flight. To analyse such a sequence,
we consider a single iterate that starts immediately
prior to the $n$th impact, as defined in the previous 
section, and ends immediately 
prior to the $(n+1)$st impact. This defines a 
\textit{bounce mapping} 
\begin{equation}
g: \:  
\begin{pmatrix}
u_n^- \\ v_n^- 
\end{pmatrix}
\mapsto_{\mbox{impact map $i$}} 
\begin{pmatrix}
u_n^+ \\ v_n^+ 
\end{pmatrix}
\mapsto_{\mbox{free flight map $f$}} 
\begin{pmatrix}
u_{n+1}^- \\ v_{n+1}^-
\end{pmatrix}  .
\end{equation}

Now, it is easy to see that for chatter to occur at all we need the free normal
acceleration to be towards the contact, that is $b<0$. Moreover, we assume
that the impact we are analysing is sufficiently far into the sequence that
the normal velocities are small. Then the
time spent in free flight is 
\beq
t_n=- 2(v_n^+/b) + O[(v_n^+)^2].
\label{eq:tn}
\eeq
Therefore we can write the 
free flight map, up to order $(v_n^+)^2$ as
$$
f:  
\begin{pmatrix}
u_n^+ \\ v_n^- 
\end{pmatrix}
\mapsto
\begin{pmatrix}
u_n^+ - (2a/b) v_n^+ \\ - v_n^+ 
\end{pmatrix},
$$
and define an effective normal coefficient of restitution via
the ratio
\begin{equation}
e := \frac{v_{n+1}^-}{v_n^-}. 
\label{eq:edef}
\end{equation}
Note that because $v_{n+1}^- = -v_n^{+}$ this ratio $e$ is precisely the Newtonian
coefficient of restitution, which ignores the motion in the tangential 
direction.

If $e<1$ then we have a chattering sequence that accumulates as 
$n \to \infty$, and the times of flight 
$t_n$ represent a geometric sequence, whose sum converges to a finite
limit which is proportional to $(1-e)^{-1}$. In the context of
impact oscillators this process is
sometimes called a \textit{complete chattering sequence} 
(see \cite[Ch.6]{Springerbook} and references therein); in the context of 
hybrid systems, the limit point of this process is sometimes called
a \textit{Zeno point} (see e.g.~\cite{Ames}). The process is like that of 
a bouncing ball coming to rest in finite time. For non-oblique impacts
(i.e.~when $B=0$) it is possible to show that $e=r$, 
a property that still holds true whenever there are no transitions from
slip to stick during the impact process 
(in regions 1,2 of Fig.~\ref{fig:impactmap}).

However, Nordmark \textit{et al} \cite{paper2} show that 
under situations where there is a transition from slip to stick 
during impact, then it is possible
to find  configurations such that $e>1$, even if the energetic
coefficient $r<1$. This seems paradoxical. If the impact process
itself cannot gain energy, then how can a chattering sequence 
occur such that the normal velocity increases from infinitesimal
values to finite ones? The resolution is that the bounce
map in this case represents a process by which energy is being
scavenged from the tangential degree of freedom and transferred to
the normal degree of freedom, such that total energy is 
still dissipated by a factor $r_e^2$ in each bounce. 

\begin{figure}
\begin{center}
\includegraphics[width=0.8\textwidth]{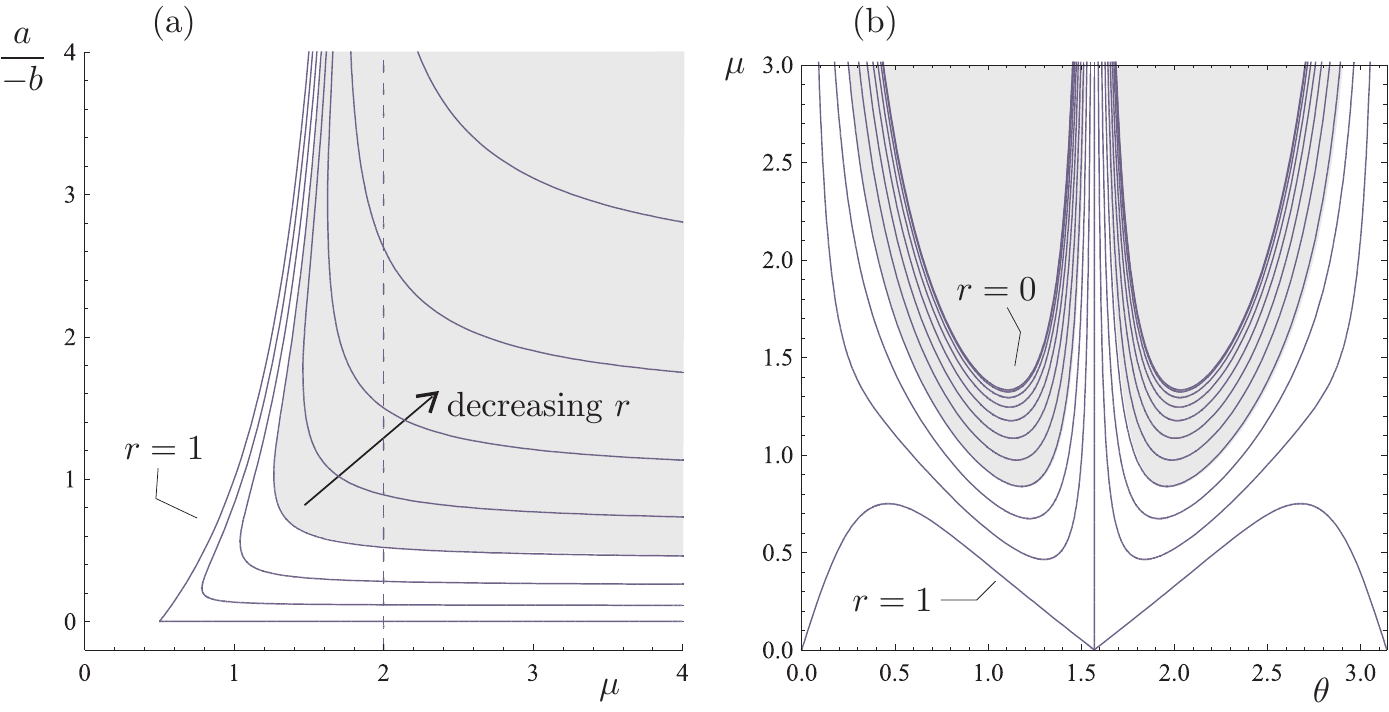}

\includegraphics[width=0.8\textwidth]{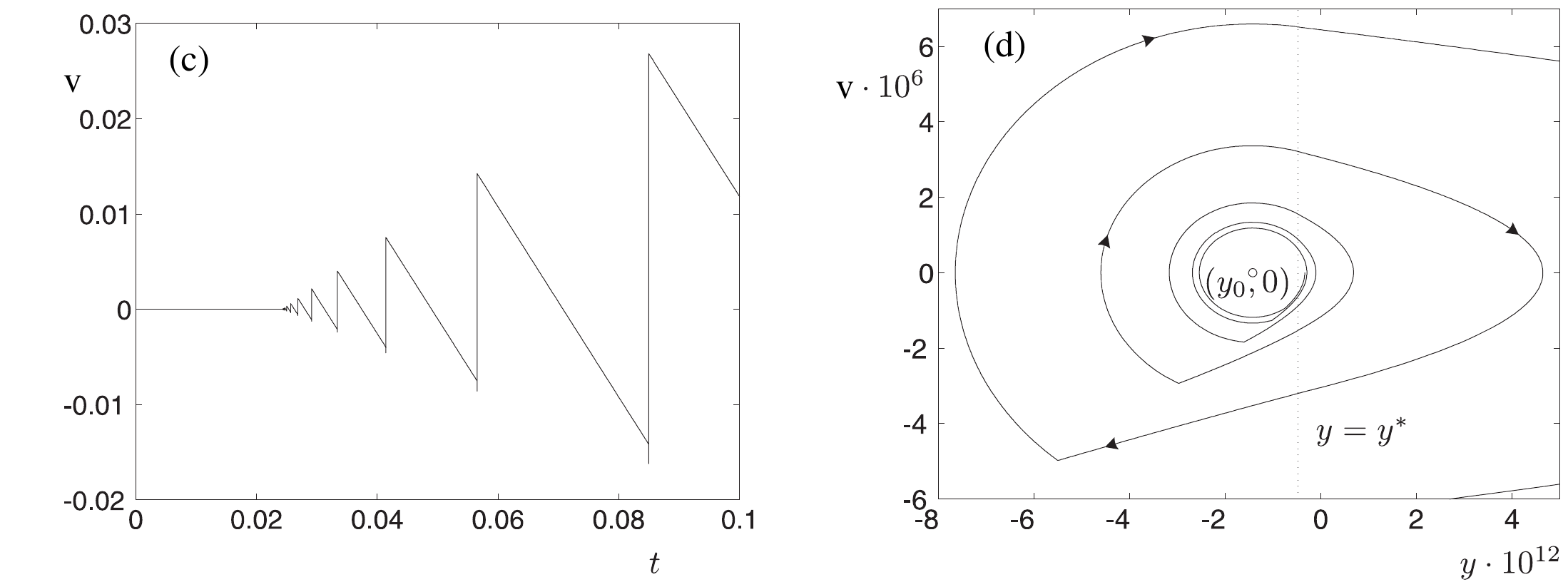}
\end{center}
\caption{Reproduced from \cite{paper2}, with permission.
(a) The region in $a/(-b)$ and $\mu$, for fixed $B=0.5$, $C=A=1$ and various
$r$-values, for which $e>1$. The region for
which reverse chatter is possible is to the right of the curve depicted
for each $r$-value. The case $r=0.7$ leads to the shaded region. 
Note in this case that the \pain region $p^+<0$ is given by $\mu>2$.
(b) The same plot for the specific case of the CPP \eqref{eq:CPP}
in terms of the angle $\theta$. Here the region above each curve for
fixed $r$ represents where there exists a range of $\dot{\theta}$-values
for which reverse chatter is possible. 
(c),(d) Simulation of a reverse chatter event in the equations for a falling rod
with applied body forces, using a stiff compliant model in
which the penetration $y$ is allowed to be violated as in 
\eqref{eq:stiffmodel}, see \cite{paper2} for the details.
}
\label{fig:chatter}
\end{figure}

The case $e>1$ gives the possibility of \textit{reverse chatter} \cite{paper2}. 
That is,
where the sequence defined by the bounce mapping $g$ converges as
$n \to -\infty$. Note 
that such a situation would lead to an extreme
form of indeterminacy. We would have an infinite sequence of increasing bounces
that starts at some time $t_0$ with zero displacement and an arbitrary phase. This arbitrary
constant gets multiplied by a factor $e>1$ at each bounce, so that at
an $O(1)$-time later there are a continuum of different possible
solutions, separated in phase by the finite time interval $t_n$, given by
\eqref{eq:tn}. Since $t_n \to 0$ as $n \to -\infty$ in this case,
we find that trajectories for all these different phases
emerge from the same initial condition. 
Fig.~\ref{fig:chatter}(c),(d) illustrates the phenomenon. 

The paper \cite{paper2} enumerates carefully the cases for which
reverse chatter is possible; the results are summarised in
Fig.~\ref{fig:chatter}(a),(b). Note that it is not a necessary
condition that we are in the \pain region. In particular, panel
(b) illustrates the conditions for $e>1$ for the CPP, and here
reverse chatter can be triggered all the way down to $\mu=0$
in the case of perfectly elastic collisions $r=1$.

\subsection{Stability}
\label{sec:3.4}

Most forms of motion within rigid-body systems with contact 
can be described by special solutions
(e.g.~equilibria, limit cycles or invariant sets) of an appropriately defined
dynamical system. In order to understand which of these motions 
may be observed in practice, it is common to examine the stability
of these solutions.
There are however several notions of stability in dynamical systems, 
and the choice of an appropriate definition is made more
subtle by the presence of unilateral constraints. 
Fundamentally though, to define stability we really need to
understand two things; what kind of motion is being considered as
being stable, and what kind of perturbations should we be stable against.

The simplest kind of motion is equilibrium. Equilibrium configurations
in contact only exist if every contact is in stick. 
We shall consider the case of multiple contact points in Sec.~\ref{sec:6}. 
For the present we shall confine
ourselves to configurations with a single point of contact. 
One characterisation of stability of an equilibrium in stick contact is
that the the system should resist small variations of 
external forces.
The difficulty of such a definition is that such forces are not usually 
considered as states of the system, rather as external inputs or parameters.
Qualitative invariance of the state of a dynamical system under changes to 
inputs or parameters is sometimes referred to as robustness (or roughness)
of the system, or more general to structural stability rather than dynamic
stability. Stability to changes in external forces is sometimes therefore
studied by contact regularisation, in which a (large) finite contact 
stiffness is introduced, so that a change in force 
necessarily implies a change in the system state. 
Contact regularisation in the normal direction forms the
subject of Sec.~\ref{sec:4} below. 

More generally, for systems with rigid contact, 
we can distinguish between \textit{normal stability}
and \textit{tangential stability}, depending on whether we allow state
perturbations at contact points that have components in the normal
direction or not. Normal stability requires stability against the
possibility of lift-off or impact, whereas tangential stability
presumes that contact is always maintained.  The study of the former
generally requires normal contact regularisation, whereas tangent
stability can be conducted using Filippov theory \cite{Filippov},
without the need to introduce extra compliance.

The key question in tangential stability is whether general motion in stick 
(not just restricted to equilibria) can
be unstable to perturbations that would cause slip. This
question is considered in 
\cite[Sec.~3]{paper2} 
for the 2D single-contact case. 
It is shown that 
provided $p^{\pm}>0$ and the contact
is in interior of the friction cone, then stick (with $u=0$) represents 
a stable \textit{sliding manifold} in the sense of Filippov (see also
\cite{Springerbook}). Thus, motion in stick in case II(iv) of Table
\ref{tab:summary} is
stable to perturbations with $u\neq 0$ except at the boundary of the friction
cone where a straightforward transition into either positive or
negative slip occurs (depending on which 
edge of the friction cone is in question). 
A more subtle argument needs to be used in the case that
stick occurs in one of the \pain parameter regions (in cells II(ii),(iii),(v) or (vi)). In \cite[Sec.~3]{paper2} it is shown that if stick is
normally stable then \textit{a posteriori} it can be shown that stick is must
also be tangentially stable in the sense of Filippov. 

Upon consideration of a larger class of state perturbations than 
just small variations
of external forces, the most widely used notion of stability is
that of \textit{Lyapunov stability}, that any small perturbation should
remain bounded.  A yet stronger condition
is \textit{asymptotic stability}, namely that such perturbations
should also decay exponentially.  Equilibria involving frictional
systems rarely possess asymptotic stability, because dry friction
tends to create continuous equilibrium sets, and perturbations
typically push the system to a nearby point within the set
\cite{leine2008}.  Thus it is natural to discuss Lyapunov
stability in the context of rigid-body systems with contact. 
In the robotics community, the concept of 
\emph{strong stability} has been proposed to refer to a case where
stick is the only consistent mode with $u=0$ at each contact 
(see \cite{Pang} and Sec.~\ref{sec:6.4} below for further discussion).
Or and Rimon \cite{Or08a} refer to 
such a property as \emph{unambiguity} and show that it
is a necessary condition for Lyapunov stability of a stick equilibrium.
We propose here a generalisation of Or and Rimon's result 
to stipulate necessary and sufficient condition for 
Lyapunov stability for an equilibrium in the presence of 
any conservative external loads and a single point contact. Specifically, three 
conditions must be met:
\begin{enumerate}
\item The curvatures of the object and the contact surface must ensure 
a local minimum of the potential energy of external forces 
along trajectories within stick (the $T$ mode). If this criterion is not met, 
divergence from the equilibrium with the $T$ mode becomes possible.
\item The equilibrium must be unambiguous in the sense of
\cite{Or08a}.  Unambiguity ensures that the system may not diverge
from the equilibrium state in $F$, $S_+$ or $S_-$ modes unless it
undergoes and impact.
\item Additionally, upon defining the effective restitution 
coefficient $e$ for chatter as in \eqref{eq:edef}, then we must have 
$e<1$. 
Otherwise a small perturbation may trigger diverging reverse-chatter
motion. 
\end{enumerate}
Analogous conditions in the multi-contact case will be 
discussed in Sec. \ref{sec:6.4}

\section{Resolution of paradoxes via contact regularisation}
\label{sec:4}

Returning to Table \ref{tab:summary}, we can see that Question 1 can
be resolved in the inconsistent regions by the requirement that an
IWC must occur. This leaves the indeterminate regions.
We shall now analyse these cases in the sense of Question 3, by adopting
contact regularisation, that is, introduction of a form of finite elasticity
into the model and then passing to the limit that the stiffness
tends to infinity. There are a number of different choices
that can be made. For example, Neimark and Smirnova \cite{Neimark} propose
including both normal and tangential compliance. However, as argued
in the previous section, unlike normal stability, 
tangential stability can be analysed without the need to introduce 
compliance. 
Therefore the simplest approach, adopted here, is 
to introduce 
compliance in the normal direction only, and otherwise assume that the
assumptions of rigid body mechanics occur, including Coulomb
friction. 
A good general discussion of
normal compliance can be found in the book by Anh \cite{LeSuanAn_book}.

It is useful though to point out a promising 
alternative approach due to  
Szalai \cite{Szalai1,Szalai2}, who introduced a formulation 
that models the response of a compliant surface through a delay kernel.  
In unpublished work, Berdeni 
\cite{Berdeni} applies
this approach to the \pain paradox. By taking the fast wavespeed limit,
a regularisation occurs via additional continuous degrees of
freedom for the normal and tangential forces. 

Following Nordmark \textit{et al} \cite{paper2}, we replace
the rigid constraint $y\geq 0$ by an assumption that there can be
small $O(\epsilon)$ excursions into $y<0$, where $\epsilon>0$ is a
small parameter. Furthermore, we suppose that the normal force is
given by a specific expression
\begin{equation}
\lambda_N (\tilde{y},\tilde{v}) = 
\begin{cases} 
 0                                 & \text{for }  \tilde{y}\geq 0, \: 
\tilde{v}\geq 0 \:,  \\
f_N(\tilde{y},\tilde{v})    & \text{otherwise},
\end{cases}
\label{eq:stiffmodel}
\end{equation}
where $\tilde{y}= \epsilon^{-1}y$; $\tilde{v}=d\tilde{y}/d\tilde{t}$;
$\tilde{t}=\epsilon^{-1/2}t$. 
Note that the timescale associated with contact is
$O(\epsilon^{1/2})$, which explains the scaling of variables. We
suppose that the restoring force has the following properties: 
\begin{enumerate}
\item $f_N$ is continuous and   
$f_N(0,0)=0$ so that $\lambda_N$ is also continuous, and
$f_N \to 0$ as $\tilde{y} \to + \infty$ 
\item $f_N$ is also smooth (at least of class $C^1$) whenever $f_N>0$,
\item $f_N$ is restoring, that is 
$$ 
\frac{\partial f_N}{\partial\tilde{y}} <0 \quad \mbox{for} \quad 
f_N>0,
$$
\item $f_N$ is dissipative 
$$
\frac{\partial f_N}{\partial \tilde{v}} < 0 \quad \mbox{for} \quad  f_N>0. 
$$
\end{enumerate}
Such assumptions are essentially equivalent to replacing the normal
rigid contact with a (possibly nonlinear) spring and damper in parallel 
that reach the rigid limit as $\epsilon \to 0$. Moreover, it is also
possible to choose the specific function $f_N$ such that it relaxes to
a (Poisson or energetic) impact law as $\epsilon \to 0$.

\subsection{Normal stability of free fall}
\label{sec:4.0}

The free fall mode does not involve active contacts, i.e.~$\lambda_N=\lambda_T=0$. The fact that the contact forces are zero makes them 
robust against small perturbations. That is if $y>0$ then small perturbations
in normal force cannot affect the motion. Alternatively, free motion
can occur if $y=0$ and $\dot{v}>0$. Here small perturbation of normal force
will not be sufficient to make $\dot{v}=0$. 
Hence the system remains in $F$ mode in response to 
small perturbation, which means that we can consider the $F$ mode to be
normally stable.

\subsection{Normal stability of stick}
\label{sec:4.1}

Introducing the  definition \eqref{eq:stiffmodel}
into the formulation
\eqref{eq:udynamics}, \eqref{eq:vdynamics} 
and assuming the conditions for stick apply, then we have
an equilibrium solution in the $y$-direction for which, according to 
\eqref{eq:stickforces},
$$
\lambda_N = \lambda_N^0 = \frac{aB-Ab}{AC-B^2} >0  \label{eq:stickforces}
$$
and hence we have an equilibrium penetration 
$\tilde y_0<0$ corresponding to stick. Now let
$$
\tilde y= \tilde y_0+\hat{y}, \qquad 
\lambda_N = \lambda^0_N + \hat{\lambda}_N := 
f_N(\tilde{y}_0,0) + k \hat{y} + q \tilde{v} +o(\hat{y},\tilde{v}),
$$
where 
$$
k = \left . \frac{\partial f_N}{\partial \tilde{y}} \right |_{(\tilde y_0,0)} \qquad
q = \left . \frac{\partial f_N}{\partial \tilde{v}} \right |_{(\tilde y_0,0)}.
$$
are well defined, by property 3 above. Substituting these expressions into
\eqref{eq:udynamics} and \eqref{eq:vdynamics}, to leading order we obtain
$$
\frac{d}{d\tilde t}
\left[\begin{array}{c}
\hat{y} \\ \tilde v
\end{array}\right]
=
\left[\begin{array}{cc}
0 & 1 \\
K k
& 
K q
\end{array}\right]
\left[\begin{array}{c}
\hat{y} \\ \tilde v
\end{array}\right], 
$$
where 
$$
K= \frac{aB-Ab}{AC-B^2}  >0.
$$
Note, from properties 1 to 4 above, that $k<0$ and $q<0$.
Hence the $2\times 2$ matrix has two eigenvalues with negative
real part. We conclude that
the equilibrium $\tilde{y}=y_0$ is normally stable. 
Taking the limit $\epsilon \to 0$, we conclude that stick is normally stable.
Note that although $y_0 \to 0$ in this limit, the normal force $\lambda^0_N$ 
remains finite. So in force space, the $T$ mode is not close to the $F$ mode
and so stick remains stable against small perturbations. However, the 
asymptotic stability occurs on the fast time-scale $\tilde t$ and
so as we pass to the limit $\epsilon \to 0$, 
the normal Lyapunov exponent tends to $-\infty$. 

 Thus, upon the introduction of compliance and taking
the infinite stiffness limit, a firm conclusion can be made within
cells II(ii) and II(iii) of Table \ref{tab:summary}. That is, if a
body is in the state of sustained stick, then this motion is stable to
small perturbations in forces  and the only way to leave sticking is by reaching
the boundary of the friction cone. In particular, a body that is in
contact and sticking will remain in stick, even if the \pain region
with $b>0$ is reached, provided that the body remains inside the
friction cone.

\subsection{Normal stability of slip}
\label{sec:4.2}

Without loss
of generality, consider positive slip.  Proceeding similarly to above,
we end up with an equation for the normal dynamics that reads
$$
\frac{d}{d \tilde{t}}
\left[\begin{array}{c}
\hat{y} \\ \tilde{v}
\end{array}\right]
=
\left[\begin{array}{cc}
0 & 1 \\
p^+ k
& 
p^+ q
\end{array}\right]
\left[\begin{array}{c}
\hat{y} \\ \tilde{v}
\end{array}
\right], 
$$
where $k,q <0$. Hence there are two eigenvalues with negative real part 
if  $p^+>0$ and two real eigenvalues of opposite sign if $p^+<0$. 
Thus we conclude
that the normal stability of the equilibrium in the fast-scale dynamics is
determined by the sign of $p^+$
Thus positive slip is stable to perturbations in the normal direction 
if and only if the appropriate \pain parameter $p^+$ is positive. 
Hence forward slip is unstable in the \pain
region $p^+<0$ (cell I(iii) in Table \ref{tab:summary}). 
A similar conclusion holds for  negative slip in the case $p^-<0$ 
(cell III(iii)). 

Thus we have that no configuration can
smoothly evolve during slip into the \pain region with $b>0$, as this
state would be violently unstable, with a normal Liapunov exponent equal
to $+\infty$ in the limit $\epsilon \to 0$. That is, it would be 
like trying to find an initial condition
that can make an infinitely heavy pin balance on its point. 
This essentially provides a unique answer
to Question 2. 
However, there is a sting in the tail, 
because if a configuration satisfied the conditions to be in such a 
`Painlev\'{e} slip' case at time $t=0$, it must leave, because the state is violently unstable. However, thinking of the compliant limit again with 
finite $\epsilon$, there are two possibilities.  
From the unstable equilibrium depth $\tilde y_0<0$ the body could
perhaps ``fall'' into a state  with $f_N<f_N^0$ in which case it must lift off
into free motion.  However it could also ``fall'' to the other side, 
$f_N>f_N^0$, in
which case an IWC would occur. After the IWC has terminated, the body would
have very different values of tangential and normal velocities than if it
had just lifted off. So there remains an indeterminacy, in the sense of
Question 1 of the introduction. An indeterminacy that is, not between
slip and lift-off but between lift-off and IWC.  It becomes imperative therefore to understand how a system can transition into a state of painlev\'{e} slip. 

\subsection{Updated contact mode consistency}

\begin{table}
\begin{center}
\begin{tabulary}{\textwidth}{|R|J|J|J|}
\hline
   &I: \: $u>0$ & II: \: $u=0$ & III: \:  $u<0$ \\
   \hline
   (i): $b>0$, $p^+>0$, $p^->0$  &$F$&$F$&$F$\\
    (ii): $b>0$, $p^+>0$, $p^-<0$  & $F$ & $F$  or  ($F$ and $T$) 
& $F$ and IWC  \\
     (iii): $b>0$, $p^+<0$, $p^->0$  & $F$ and IWC   & 
$F$ or ($F$ and $T$) & $F$ \\
     \hline
  (iv): $b<0$, $p^+>0$, $p^->0$    & $S_+$ & $T$ or $S_+$ or $S_-$ & $S_-$ \\
    (v): $b<0$, $p^+>0$, $p^-<0$  & $S_+$ & $S_+$ or $T$ & IWC\\
     (vi): $b<0$, $p^+<0$, $p^->0$ & IWC & $S_-$ or $T$ & $S_-$ \\
     \hline
\end{tabulary}
\caption{Updated version of Table \ref{tab:summary}, taking into account the
results of Secs.~\ref{sec:3} and \ref{sec:4}.}
\label{tab:summary2}
\end{center}
\end{table}

\begin{figure}
\begin{center}
\includegraphics[width=0.85\textwidth]{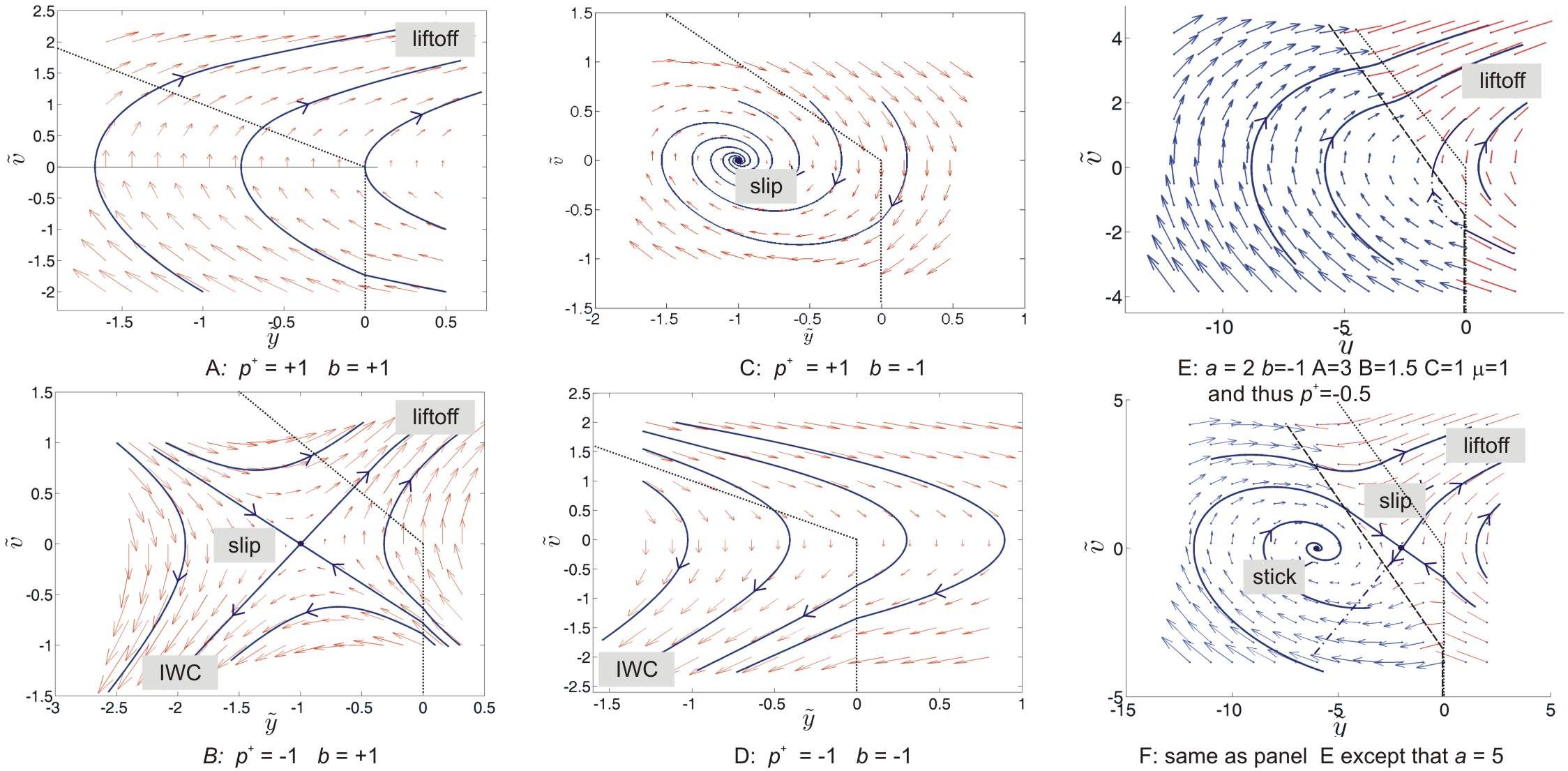}
\end{center}
\caption{Phase portraits of fast contact dynamics of a point contact with 
$k=q=-1$. Panels A--D illustrate cases for $u>0$, specifically A: case I(i-ii)
of Table \ref{tab:summary2}, B: case I(iii), C: case I(iv-v) and D: case I(vi) (panel D) of Table \ref{tab:summary2}. Panels
E and F show two different phase portraits for zero initial tangential velocity $u=0$, in case II(ii). E shows the case where lift-off always occurs, 
whereas F illustrates the case of indeterminacy between $F$ and IWC. 
Dotted lines in all panels show
the border between active ($f_N>0$) and passive ($f_N=0$) contacts. 
The dashed line in panels E and F depicts states for which the 
contact force is at the boundary of the friction cone. 
To the right of this line, there is positive tangential acceleration at
the contact point.}
\label{fig:phaseportraits}
\end{figure}

In the light of the above analysis, we can update
Table \ref{tab:summary}, to take account of the normal stability results
and the possibility of impact;  see Table
\ref{tab:summary2}.  The above-mentioned indeterminacy between lift-off 
and impact occurs in cells III(ii) and I(iii). There is also another 
indeterminacy in cells II(ii) and II(iii) where either stick or lift 
off can occur, which we shall explain shortly. 

To understand either of these indeterminancies though it is useful to 
illustrate phase portraits of the regularised contact dynamics. See
Fig.~\ref{fig:phaseportraits}, in which we have used 
\eqref{eq:stiffmodel} with 
\begin{equation}
f_N=
\max
\left[
\begin{array}{c}
k\tilde{y}+q\tilde{v}
\\
0
\end{array}
\right]
\label{eq:linearcontact}
\end{equation}
where $k$ and $q$ are negative scalars.  

Consider first cases where the initial condition is in slip and, 
without loss of generality, let us assume $u>0$. The four possible cases are 
illustrated in Fig.~\ref{fig:phaseportraits}(A-D). 
If $b<0<p^+$ (case I(iv)) then the normal dynamics has a 
globally attractive equilibrium corresponding
to positive slip. In contrast, if $b>0>p^+$ (case I(iii)) the 
positive slip normal equilibrium corresponds to a saddle point, 
and trajectories converge to $\tilde{y}=+\infty$ (liftoff) 
or $\tilde{y}=-\infty$ (IWC) depending on initial conditions. 
In the remaining two cases, slipping was found
inconsistent in Sec.~\ref{sec:3}. Accordingly the dynamics of the compliant
contacts has no equilibrium and every trajectory converges to
$\tilde{y}=+\infty$ if $0<b,p^+$, or to $\tilde{y}=-\infty$ if
$0>b,p^+$. This picture confirms the result that the contact undergoes
liftoff in case I(i) and IWC in case I(iv). 

Consider now initial conditions for which $u=0$. Then in all cases
other than II(ii-iii), phase portraits like
Fig.~\ref{fig:phaseportraits}(A) or (C) occur where now the normal
equilibrium corresponds to stick rather than slip if the equilibrium
is in the interior of the friction cone.
Fig.~\ref{fig:phaseportraits}(E) and (F) illustrate the possibly
indeterminate case II(iii) (case II(ii) is similar).  Panel E
represents the case where the configuration (ignoring normal
compliance) is outside of the friction cone and the only possibility
is motion in $F$ mode.  Here a dashed line shows those states for
which zero tangential acceleration requires $\lambda_T= - \mu
\lambda_N$. To the left of this line, the contact remains in stick,
whereas to the right of this line it has positive tangential
acceleration so that slip or lift-off must occur. Crossing from left
to right corresponds to a stick-slip transition, whereas crossing from
right to left does not represent a transition because $u>0$. These
latter trajectories do not follow the depicted (sticking) vector field
as long as the tangential velocity of the contact point remains
positive; one such trajectory is depicted by a dot-dashed
line. Nevertheless we see in this case that all initial conditions
eventually lead to lift-off, and so $F$ remains the only mode.

Fig.~\ref{fig:phaseportraits}(F) illustrates the other branch of the
``or'' in cell II(iii) of Table \ref{tab:summary} in which there are
additional feasible $T$ and $S_+$ modes.  Here we see that the $T$
mode represents a stable focus whereas slipping $S_+$ can be seen to
be a saddle point, and hence unstable, in accordance with the
calculations in Secs.~ 4.2 and 3.3 above. Note though that there is
still indeterminacy in this case, because lift off (mode $F$) is
another possibility, which is represented here by trajectories
diverging to positive infinity in $y$. Note that the stable manifold
of the saddle separates trajectories that are attracted to mode $T$
from those which lift-off.

\section{Triggered transitions between states}
\label{sec:5}
%
%

In this section we focus on Question 2, namely what are the generic ways
in which a configuration can enter a \pain region and what must then happen.
Theoretically, there are only two ways of reaching sustained contact in 
a \pain state; (1) moving from non-\pain contact to \pain contact, or 
(2) establishing sustained contact via the termination of a chattering 
sequence, inside the \pain regime. We deal with these two possibilities in
the next two subsections. We shall then look at ways in which reverse chatter
may be triggered while in contact.

\subsection{Entering a \pain region within contact; the G-spot}
\label{sec:5.1}

Suppose at some time $t_0$ that a rigid-body system 
is contact with both $p^\pm>0$. 
We want to know how to continue the dynamics for $t>t_0$. Note that 
parameters $p^{\pm}$, $a$ and $b$ will be smooth functions
of time provided the body doesn't change its 
mode of motion (between stick, slip or free motion). 
The analysis in the previous section shows that 
any body in stick should remain in stick provided it remains in
the friction cone. In particular a change of sign of $p^\pm$ would not
affect the mode of motion. This is not true of slip though. As we have seen,
if the appropriate \pain parameter becomes negative, then 
slip changes from being highly stable to highly unstable. 

For definiteness let us consider a positive slip $u>0$ within region
I(iv) of Table \ref{tab:summary2}; the analysis for negative slip is
similar. We are interested in where $p^+$ being close to zero for $b<0$. So
suppose we have an initial condition with
$0<p^+ \ll 1$. Hence, according to \eqref{eq:udotslip+}
the contact force $\lambda_N$ is large. In particular,
we can see that if $p^+ \to 0$ while $b$ remained finite, then 
$\lambda_N \to +\infty$ which implies $\lambda_T =-\mu \lambda_N \to -\infty$. 
Large contact force means large accelerations. When viewed in state
space, we find that these large accelerations must cause motion 
in the direction tangent to the  boundary $p^+=0$, and so the boundary cannot
be crossed transversely. 

In the case of systems with three degrees-of-freedom
like the CPP, and any system composed of a single rigid body with a
unique point contact, then the 
set $p^+=0$ is a 1-dimensional sub-manifold of the set 
of slipping trajectories.
A more precise analysis can be performed by realising that the above
arguments show the dynamics of slip becomes singular as $p^+ \to 0$ and
rescaling time appropriately.
Following G\'{e}not and Brogliato 
\cite{Genot1999} it then can be shown that $p^+=0$ is composed of a solution 
trajectory in this rescaled time. Thus, there can be no
crossing into the \pain region except possibly at a singular point
of the rescaled slipping vector field.  The only such 
singular point of in this case is where $b=p^+=0$. 

The analysis in \cite{Genot1999}  was specific to
the CPP; we sketch here a generalisation of what must occur for
any system close to the so-called G-spot, $b=p=0$, 
a complete version of which
will appear elsewhere \cite{paper3}.
The trick in
\cite{Genot1999} is to analysis the problem by rescaling
time 
\begin{equation}
dt=p^+ ds \label{eq:time_rescale},
\end{equation}
and then to think of the scalar variables $p^+$ and $b$ as dynamical
quantities that evolve on the timescale $s$.  From 
\eqref{eq:udynamics}, \eqref{eq:vdynamics} we then obtain, 
to leading order in $p^+$ and $b$, 
\begin{eqnarray}
\frac{d}{ds} p^+  &=& \alpha _{1} p^+  , \label{eq:Gspot1} \\
\frac{d}{ds}b &=&\alpha _{2} p^+ +\alpha _{3}b, \label{eq:Gspot2} 
\end{eqnarray}
where
\begin{eqnarray*}
\alpha _{1} &=&\dot{q} \cdot \frac{\partial }{\partial q} p^+
+\frac{\partial }{\partial t} p^+ , \\
\alpha _{2} &=& M^{-1} f \frac{\partial b}{\partial \dot{q}}+
 \frac{\partial b}{\partial q} \dot{q}+\frac{\partial b}{\partial t}, \\
\alpha _{3} &=&
M^{-1}  \left( \mu c -d \right) 
\frac{ \partial b}{\partial \dot{q}}.
\end{eqnarray*}
Here all functions are evaluated at the codimension-two point where 
$p^+=b=0$, in which case $\alpha_{1--3}$ are just real constants. 
Specifically $\alpha_{1}$ is just the time derivative 
of $p^+$ and $\alpha_{2}$ the time derivative of $b$, whereas $\alpha_3$ is
the time derivative of the dependence of $b$ on the parameter $p$. 

As an example consider the CPP for a uniform rod with $m=1$, $l=2$. Then
\begin{eqnarray*}
  \alpha _{1} &=&-3\omega \left( \sin 2\theta +\mu \cos 2\theta \right)  \\
  \alpha _{2} &=& 9R\omega \sin \theta +\omega^{3}\cos \theta+
  \dot{S}_y -3\dot{R}\cos(\theta)\\
  \alpha _{3} &=&-6\omega \sin \theta \left( \mu \sin \theta -\cos \theta
  \right), 
\end{eqnarray*}
where $S_y$ and $R$ are the $y$-component of torque component of 
body forces at the centre of mass. 
Note that for general configurations, such as those reviewed in Sec.~2,
the $\alpha$ parameters can take any combination of signs. 

\begin{figure}
\begin{center}
\includegraphics[width=0.9\textwidth]{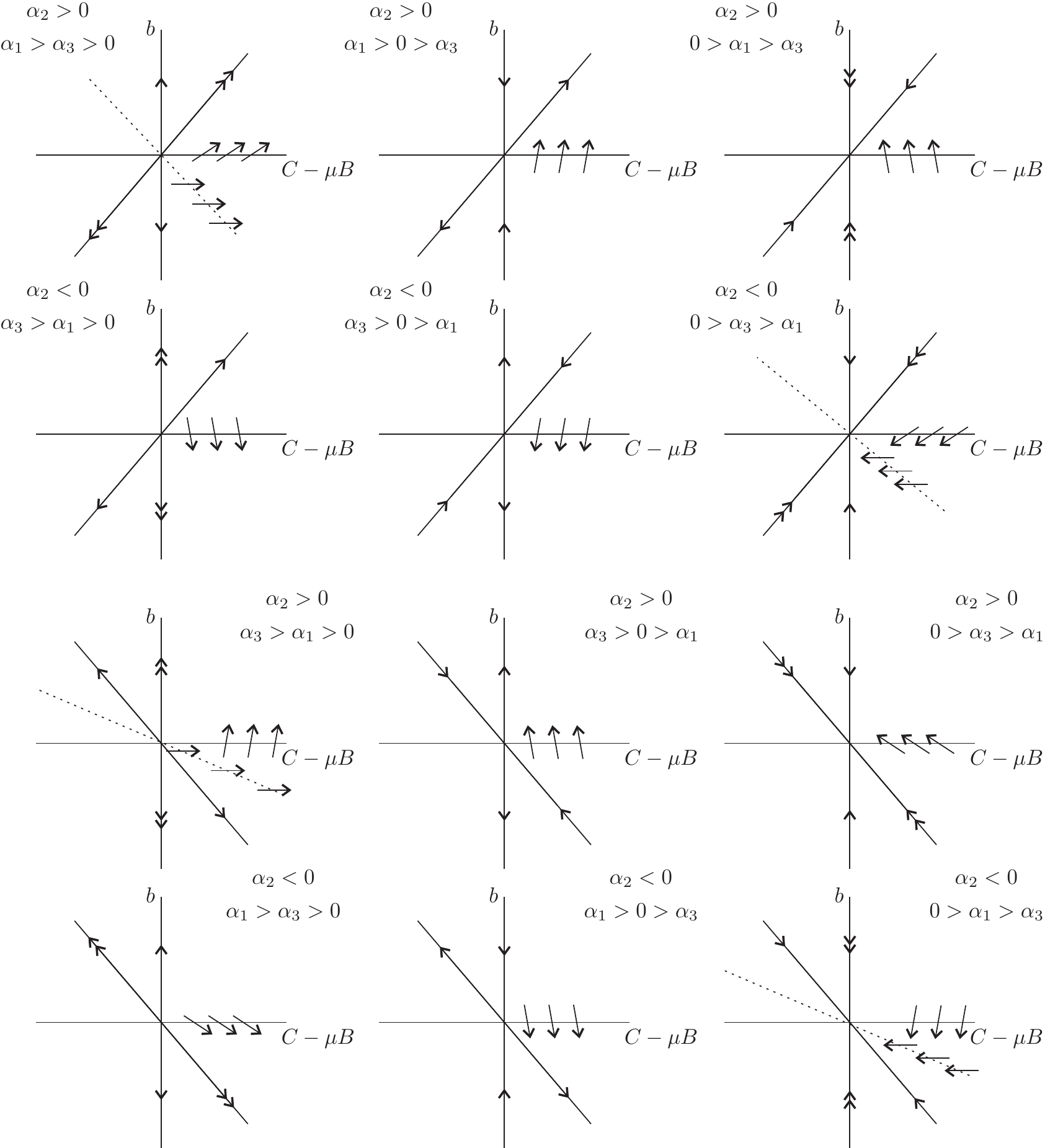}
\end{center}
\caption{Reproduced from \cite{paper3}.  Analysis of the dynamics of
  the neighbourhood of the G-spot.  In each panel, the phase plane of
  \eqref{eq:Gspot1}, \eqref{eq:Gspot2} is depicted, for all possible
  sign combinations of the parameters $\alpha_{1,2,3}$. Here the bold
  line passing through the origin represents the location of the
  second eigenvector, and the arrows on this line and the $b$-axis
  represent whether each represents a stable or unstable manifold. In
  the case of a two-dimensional stable (unstable) eigenspace, double
  arrows are used to represent the strong stable (unstable)
  manifold. The dashed line represents the location of the horizontal
  nullcline. Arrows crossing this line and also the $(C-\mu B)$-axis
  indicate the direction of solution trajectories there }
\label{fig:Gspot} 
\end{figure}

Note that in the rescaled singular timescale $s$, the G-spot is an equilibrium
point and the dynamical system \eqref{eq:Gspot1}, \eqref{eq:Gspot2} provides a 
planar linear system whose dynamics govern the behaviour near this singular
equilibrium. Note that the singularity of the time rescaling at the G-spot
means that the system \eqref{eq:Gspot1} \eqref{eq:Gspot2} only makes sense
in the original co-ordinates if $p^+ \neq 0$. 

Figure \ref{fig:Gspot} shows all possible 
qualitatively distinct phase portraits close to the G-spot.
We are interested in what happens to initial conditions that are initially
in slip, that is in the bottom right quadrant $p^+>0$, $b<0$. 
To explain the figure, note that the coefficient matrix 
$$
\begin{pmatrix}
\alpha_1 & 0 \\
\alpha_2 & \alpha_3
\end{pmatrix}
$$
has two eigenvalues $\alpha_{1}$ and $\alpha_3$ with eigenvectors
$(\alpha_{1}-\alpha _{3}, \alpha _{2})^T$ and
$(0,1)^T$, respectively. In particular, since it always corresponds to
an eigenvector, independently of the value the coefficients $\alpha_{1,2,3}$, 
the $b^+$ axis is invariant.
This latter fact shows immediately that an initial condition
with $p^+>0$, $b<0$, cannot evolve into the \pain region $p^+<0$ except
possibly at the G-spot $p^+=b=0$. In fact Fig.~\ref{fig:Gspot}
shows that there are only three possible outcomes for any open set of
initial conditions starting in the bottom right quadrant: 
\begin{enumerate}
\item {\bf lift off} via passage through $b=0$ for $p^+>0$,
\item {\bf continued slip} by moving away from a neighbourhood of the origin
while remaining in the bottom right quadrant,
\item {\bf approaching the G-spot}. 
\end{enumerate}
The first two possibilities are regular transitions that do not
involve the singularity $p^+ \to 0$. Possibility 3 is rather special though,
and although the G-spot is an isolated codimension-two point in phase space,
it can attract an open set of initial conditions, owing to the fact that
it is an equilibrium point in the rescaled dynamics. 

Note that this third possibility can only occur in panels (c), (e) and (i) of
Fig.~\ref{fig:Gspot}, which really split into two separate cases. 
Either convergence to the
G-spot is tangent to the $b$ axis, or it is tangent to the 
$\alpha_1$-eigenvector that is the line $\alpha_2 p^+ = (\alpha_1-\alpha_3) b$.
Note in the former case we have that  
$\lambda_N = -b/p^+ $ 
becomes infinite, as we approach the G-spot, whereas in the former case
$\lambda_N$ tends to the finite limit associated with the non-trivial
eigenvector. Recalling that in the unscaled time,
the G-spot is reached in finite time, we thus have two possibilities:
\begin{description}
\item[G1] If $0>\alpha_3>\alpha_1$ and the initial condition 
$(p^+,b)$ in the lower right quadrant close to the G-spot is such that 
$\alpha_2 p^+ < (\alpha_1-\alpha_3 ) b $, then the G-spot is approached in
finite time such that ratio $p^+/b \to 0$ and the 
normal force $\lambda_N \to \infty$. 
\item[G2] If $\alpha_2<0$, $0<\alpha_1<\alpha_3$ then the G-spot is
approached in finite time, and $\lambda_N$ approaches the 
finite limit $\alpha_2/(\alpha_3-\alpha_1)>0$. 
\end{description}

Each of these cases correspond to what has been referred to in the literature
as \textit{dynamic jam}. What happens next beyond dynamic jam 
remains an open question. In principle after the G-spot, there could
be an IWC, a lift off event or even possibly a velocity jump immediately
into stick (recall that $u>0$ as we approach the G-spot). 
It seems that these possibilities cannot easily be distinguished using
the contact regularisation approach of Sec.~\ref{sec:4}.

\subsection{Reaching a \pain state via complete chatter}
\label{sec:5.2}

The analysis in Secs.~\ref{sec:3} and \ref{sec:4} analyse both stick
and slip in the \pain regions, assuming that a trajectory arrived in
the \pain state by smooth evolution while in contact. We found that
stick is unaffected by either $p^\pm<0$, provided the body is in
the interior of the friction cone.  In contrast, the analysis in
Sec.~\ref{sec:5.1} above shows that the only way to enter the \pain
region while slipping is via the G-spot. What we have not investigated
though is whether a complete chatter sequence can end up a Zeno point
that is in the \pain region.

In order for a chattering sequence to occur, we must have $b<0$. Hence
the indeterminate cases of stick (cells II(ii) and III(ii) in Table
\ref{tab:summary2}) cannot be reached. So, consider instead the possibility 
of slip.  Suppose that the parameter
$p^+<0$ and consider the possibility of a chattering sequence leading
to positive slip.  Here, we are in the case of impact mappings as
depicted in Fig.~\ref{fig:impactmap}(b). Note that any impact with an
initial velocity with $u^->0$, must end up with $u^+=0$. Similarly any
impact with $u^-<0$ ends up with $u^+\leq 0$. Hence the only
possibility is that the Zeno point has $u\leq 0$. But note that there
is no inconsistency or indeterminacy in this case (see cells I(vi) and
II(vi) in Table \ref{tab:summary2}). We either get stick or negative
slip, the latter of which is consistent because $p^->0$.

A similar conclusion holds if $p^-<0$, we cannot approach the \pain
region for negative slip as the result of a chattering sequence.
We conclude that even if one of the parameters $p^\pm<0$, it is not
possible to have a Zeno point in the \pain region, thus dismissing this is
a possible route into the \pain region. 

\subsection{Triggering reverse chatter}

A final possibility for a complex triggered transition would be if a
body that is smoothly in contact, were to spontaneously undergo a reverse
chatter sequence. This possibility was considered by Nordmark \textit{et al}
\cite{paper2}, the results of which we summarise here. 

In general, we can rule out such a spontaneous
triggering of reverese chatter in T mode by the analysis of Sec.~\ref{sec:4.1} since, provided the body remains in the
interior of the friction cone, stick is stable. Similarly, the
analysis of Sec.~\ref{sec:4.2} rules out the possibility of
spontaneous triggering of reverse chatter from within a condition of
sustained slip, as this is stable wherever it is
consistent (unless the border of \painx -ness is crossed, 
which was shown to be impossible). 
Similarly, reverse chatter cannot be triggered at the end
of a forward chattering sequence, because such a sequence necessarily
converges to a point where $e<1$.

This leaves the possibility that reverse chatter can be triggered at
the boundary of one of the contact phases, namely at the termination
of stick, or of slip. But, an analysis of the cases where $e$ can be
greater than one shows that these are necessarily a subset of the cases
that are in the interior of the friction cone. Hence it can be concluded
that reverse chatter can only be triggered at the end of a slip phase.
Then, it can be argued that either a case of stick is triggered, or
under certain other conditions, in addition to $e>1$, we might get 
the triggering of reverse chatter.

Nordmark \textit{et al} \cite{paper2} enumerate the details and find
that there are further inequality conditions on $a$, $b$, $A$, 
$B$, $C$, $\mu$ and $r$ that determine whether reverse chatter can be triggered
from a transition into stick. Then they show, by introducing a compliant
model, as in Sec.~\ref{sec:4}, that if these conditions are satisfied then
the situation is not uniformly resolvable. 
That is, the precise details of
the compliant force $f_N$ can affect whether regular stick motion of 
reverse chatter is triggered. We are left with another indeterminate case. 

\section{Multiple contact points}
\label{sec:6}
%
%

Let us now consider generalisation of the previous results to systems
with multiple contacts. One aim is to identify which
properties of \painx's paradox are generally true, and which are
specific properties of single-contact systems. Moreover, we will
uncover many extra forms of degeneracy that are unique to
multiple-contact point systems.

One of the main difficulties with answering Question 1 is that if
we allow $F$, $T$, $S_+$ and $S_-$ at each of $n$ 
contacts, then this induces $4^n$ possible contact modes (although some of
them will be kinematically inadmissible). This combinatorial complexity 
makes a consistency check for each potential combination 
computationally unfeasible for large $n$. In
what follows we therefore focus mostly on the simplest case of just two 
frictional contacts.

Ivanov \cite{Ivanov03} considered generalisations of the CPP for a rod
that has two points of contacts with rough rigid surfaces under the assumption
that each contact undergoes slip. 
He enumerated the conditions under which the \pain paradox appears for
suitably chosen external forces. The classification scheme of Ivanov
has recently been refined by V\'arkonyi 
\cite{Varkonyi15}
to the level of our analysis in
Sec.~\ref{sec:3.1}. A summary of and commentary on 
these results appear in the next three subsections. 

\subsection{Consistency analysis of contact modes}
\label{sec:6.1}

To begin with, consider a planar rigid-body system that has 
two point contacts, with nonzero tangential velocities: 
$x_i=y_i=v_i=0$ and $u_i>0$ for $i=1,2$. Thus there are 
four kinematically admissible modes: $S_+S_+$, $S_+F$,
$FS_+$, and $FF$. 
Here the two-letter symbols refer to the specific combination of modes
at the two contacts. The contact forces and accelerations are determined
with the help of the equality constraints of Table \ref{tab:summary}
for both contacts. A contact mode is deemed consistent, if the consistency
conditions of Table \ref{tab:summary} are satisfied for both contacts.

The key observation is that the general analytical expressions for
contact motion in Sec.~\ref{sec:3} still apply, specifically the equation
\eqref{eq:normaldyn_02_slip+}, 
but where now 
$b$ and $\lambda_N$ are vectors in $\Rset^2$ and $p^+$ is a matrix of size
$2\times2$. In this formulation, the elements $p^+_{ij}$ of $p^+$ represent the
normal acceleration of contact $i$ in response to a slipping contact
force with a normal component of unit size at contact $j$.

V\'{a}rkonyi \cite{Varkonyi15} has recently enumerated all the consistent
modes under this restriction of no sticking modes. 
In order to summarise the results of his analysis, it is convenient to 
introduce the following notation. Let $p_j$ represent 
the $j^{th}$ column vector of
$p^+$ such that $p^+=[p_1\; p_2]$. Furthermore, let $\gamma_j$ be the
angle between $p_j$ and vector $[1\;0]^T$. For example, if
$p_j=[0\;1]^T$ then $\gamma_j=\pi/2$.  Similarly, we let
$\hat\gamma_j$ be the angle between $-p_j$ and the vector $[1\;0]^T$
and $\beta$ be the angle between $-b$ and $[1\;0]^T$.  It can be shown
that the consistency of each contact mode can be determined using a
string of three integers
$$
s_1 s_2 s_3, \quad \mbox{where} \quad s_1 \in \{1,2,3,4\}, \quad 
s_{2,3} \in \{1,2,\ldots 6\},
$$
which code the relative position of the vectors 
$p_1$, $p_2$, $b$ and the co-ordinate axes. 
Specifically the index $s_1$ represents the quadrant of $\Rset^2$  that 
contains $p_1$. Equivalently, $s_1$ is the index of the interval among
the four ordered
intervals $[0,\pi/2]$, $[\pi/2, \pi]$, 
$[\pi,3\pi/2]$, $[3\pi/2,2\pi]$ that contains $\gamma_1$. In a similar 
fashion, the second integer $s_2$ represents the index of the interval among
the set of ordered closed intervals bounded by  
$0$, $\hat{\gamma_1}$, $\gamma_1$ $\pi/2$, $\pi$, $3\pi/2$ and $2\pi$ 
that contains
$\gamma_2$. The final integer $s_3$ represents the index of the 
interval among
the set of ordered closed intervals bounded by  
$0$, $\gamma_1$, $\gamma_2$ $\pi/2$, $\pi$ $3\pi/2$ and $2\pi$ that contains
$\beta$. 

\begin{figure}
\begin{center}
\includegraphics[width=15cm]{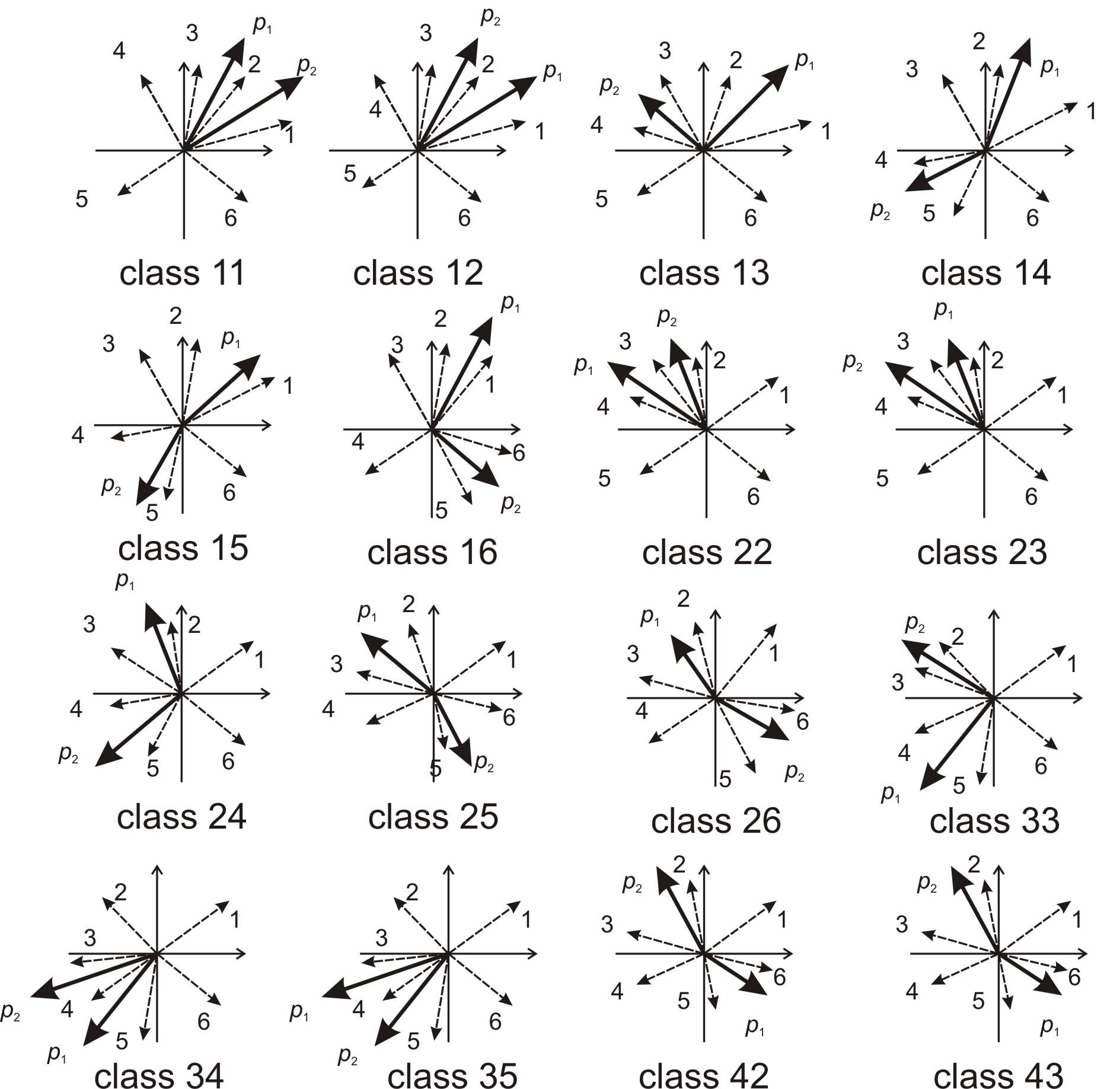}
\caption{Relative positions of the vectors $p_1$ and $p_2$ (solid arrows)
and the four co-ordinate semi-axes  
in each class $s_1s_2$. In each case, every
possible position of the third vector $b$, 
which delineates the third integer $s_3$ is indicated by a dashed arrow. Note that only one member of each dual 
pair of classes is displayed, see Table \ref{tab:classification} for
the corresponding dual $\hat{s}_1 \hat{s}_1$ in each case.}
\end{center}
\label{fig:classification}
\end{figure}

The relative positions of the vectors $p_1$, $p_2$ and $b$ 
corresponding to each class of configurations is 
depicted in Fig.~\ref{fig:classification}. 
Thus, for example, the upper right panel refers to 
classes 14$s_3$ where $s_3=1,2,3 \ldots 6$.  
Here we see, according to the above notation, that  
$s_1=1$ and so $p_1$ is in the first quadrant. In addition, 
$s_2=4$ and so $p_2$
is in third quadrant with positive angle to the $[0\; 1]^T$ axis
that is less than that of $-p_1$. The $s_3$-value corresponding to 
possible locations of
$b$ are indicated by dashed arrows, which
depicts the relative position of the vector $b$ with respect 
to $p_1$, $p_2$ and
the four co-ordinate semi-axes.  

\begin{table}[]
\centering
\begin{tabular}{| l | cccc | ccc |}
\hline
                  Classes        & \multicolumn{4}{l}{Contact modes}                                                                   & \multicolumn{3}{l}{IWC} \\ \hline
name (dual) & $FF$    & $S_+S_+$                          & $S_+F$                           & $FS_+$                         & $IF$     & $FI$    & $II$     \\ \hline
11 (self)                 & 5     & \cellcolor[HTML]{9B9B9B}2     & 1 2 6                         & 2 3 4                       &        &       &        \\
12 (self)                 & 5     & 2                             & 1 6                           & 3 4                         &        &       &        \\
13 (41)                   & 5     & \cellcolor[HTML]{F8FF00}2 3   & 1 6                           & 4                           &        &       &        \\
14 (31)                   & 4 5   & \cellcolor[HTML]{F8FF00}2 3 4 & 1 6                           & \cellcolor[HTML]{9B9B9B}4   &        &       & (x)    \\
15 (32)                   & 4 5   & \cellcolor[HTML]{9B9B9B}1 5 6 & 1 6                           & \cellcolor[HTML]{9B9B9B}4   &        &       & (x)    \\
16 (21)                   & 4     & \cellcolor[HTML]{9B9B9B}1 6   & 1 5 6                         & \cellcolor[HTML]{9B9B9B}4 5 &        & x     &        \\
22 (45)                   & 5     & \cellcolor[HTML]{9B9B9B}3     & \cellcolor[HTML]{9B9B9B}4 5   & 3 4                         & x      &       &        \\
23 (46)                   & 5     & \cellcolor[HTML]{F8FF00}3     & \cellcolor[HTML]{9B9B9B}3 4 5 & 4                           & x      &       &        \\
24 (36)                   & 4 5   & \cellcolor[HTML]{9B9B9B}3 4   & \cellcolor[HTML]{9B9B9B}3 4 5 & \cellcolor[HTML]{9B9B9B}4   & x      &       & (x)    \\
25 (self)                 & 4     & \cellcolor[HTML]{9B9B9B}3 4 5 & \cellcolor[HTML]{9B9B9B}3 4   & \cellcolor[HTML]{9B9B9B}4 5 & x      & x     & (x)    \\
26 (self)                 & 4     & \cellcolor[HTML]{9B9B9B}1 2 6 & \cellcolor[HTML]{9B9B9B}3 4   & \cellcolor[HTML]{9B9B9B}4 5 & x      & x     &        \\
33 (44)                   & 4 5   & \cellcolor[HTML]{9B9B9B}3 4   & \cellcolor[HTML]{9B9B9B}5     & \cellcolor[HTML]{FFFFFF}3   &        &       & (x)    \\
34 (self)                 & 3 4 5 & \cellcolor[HTML]{9B9B9B}4     & \cellcolor[HTML]{9B9B9B}5     & \cellcolor[HTML]{9B9B9B}3   &        &       & (x)    \\
35 (self)                 & 3 4 5 & \cellcolor[HTML]{9B9B9B}4     & \cellcolor[HTML]{9B9B9B}4 5   & \cellcolor[HTML]{9B9B9B}3 4 &        &       & (x)    \\
42 (self)                 & 4     & 1 2 6                         & 5                             & 3                           &        &       &        \\
43 (self)                 & 4     & \cellcolor[HTML]{9B9B9B}3 4 5 & 5                             & 3                           &        &       & (x)  \\ \hline 
\end{tabular}
\caption{Classification of planar systems with two non-sticking contacts. 
See text for details.} 
\label{tab:classification}
\end{table}

The results of the consistency analysis are summarised in
Table~\ref{tab:classification}.  To cut down the number of rows of the
table and sub-panels of Fig.~\ref{fig:classification}, we have used
the fact that each class has a dual that is obtained by swapping the
labels of contact 1 and contact 2. This corresponds to swapping the labels $p_1$
and $p_2$ and reflecting both vectors in the line $45^\circ$ line
in each case.  The dual class $\hat{s}_1
\hat{s}_2$ of each of the depicted cases $s_1 s_2$ is given in
brackets in the first column of Table \ref{tab:classification},
including cases that are self-dual. Note that the first two digits
$s_1 s_2$ of any class represents the coarser classification of Ivanov
\cite{Ivanov03}. The last digit $s_3$ refines the classification into
subclasses within which the consistency of each contact mode becomes
uniquely determined (as depicted in columns 2 to 5 of the table).  The
significance of the shading of these columns and the meaning of the
last three columns will be explained in the next subsection.

Note already that there are multiple occurrences of indeterminacy and
inconsistency indicated in the table. For example, within class 15, subclasses
151, 154 and 156 are indeterminate because two contact modes in each case are consistent. 
In contrast subclasses 152 and 153 are inconsistent because there is
no consistent contact mode for these cases. 
There are also several examples of triple indeterminacy, 
for example in subclass 112
where $S_+S_+$ and $S_+F$ and $FS_+$ are all consistent and of 
quadruple indeterminacy where all four possible contact
modes are consistent, for example in subclasses 254 or 354.

Before going on to look at normal stability of these modes and the 
possibility
of impact, it is worth stressing that Table \ref{tab:classification} is
only a partial classification for two contacts, because it ignores stick and
cases where negative and positive slip can simulateously occur at different contacts. 
As far as the authors know, a general classification
is not yet known, nor is a classification in cases where there are more than
two contacts.  Nevertheless, other examples of indeterminacy are known. Two
such cases were illustrated in
Fig.~\ref{fig:ambiguity}.  The left-hand panel depicts a 
case in which the static equilibrium represents the $TT$ mode which is
consistent provided there is a sufficiently large friction force at the
contacts (solid arrows in the figure show the gravitational force and
a possible choice of reaction forces).  Simultaneously, the $FF$ mode
is also consistent, in which case the block will just fall.  
The right-hand panel of Fig.~\ref{fig:ambiguity} 
illustrates a shape resting on a slope in
which both the body may be in equilibrium in mode $TT$, but for which
motion in mode $S_+F$ is also consistent.

\subsection{Contact regularisation and stability analysis} \label{sec:6.2}

To resolve the indeterminate cases, we can compute the normal 
stability of different kinds of contact via the method introduced in 
Sec.~\ref{sec:4}. That is, we introduce a contact regularisation in the normal
direction, linearise the ensuing equations of motion, perform
an eigenvalue analysis on the appropriate Jacobian matrix and then pass
to the infinite stiffness limit. 

Applying this methodology to the cases enumerated in
Table~\ref{tab:classification}, it is straightforward to see that a
contact in mode $F$ does not influence the normal stability.  Hence,
the stability analysis of the $S_+F$ and $FS_+$ contact modes is
identical to the analysis of $S_+$ in Sec.~\ref{sec:4}. Specifically,
we obtain that $S_+F$ is stable if $p_{11}>0$ and $FS_+$ is stable if
$p_{22}>0$. For the $S_+S_+$ mode, the simultaneous dynamics of two
contacts is described by a system of four first-order ODEs, and thus
normal stability analysis requires the study of eigenvalues of a
$4\times 4$ Jacobian matrix.  The Jacobian and thus the stability
properties of contact modes are independent of $b$. Hence the same
results are obtained for all subclasses within this class.

The results of the stability analysis are summarised in
Table~\ref{tab:classification}. Details of the calculations are presented in \cite{Varkonyi15}. Stability information is
indicated by shading of the appropriate cell in the table.  A dark
(grey) shading indicates definitely unstable. Lighter (yellow, colour
online) shading represents where the $S_+S_+$ mode may be stable,
depending on further conditions. Finally, no shading represents a
normally stable situation.  One interesting thing to note is that the
stability of the $S_+S_+$ mode is not uniquely determined in classes
13, 41, 14, 31, 23 and 46. The unstable regions within these
subclasses are caused by so-called \textit{mode coupling} -- a
phenomenon studied in the context of break squeal (see
e.g.~\cite{Ibrahim1994,Hoffmann}).  In these regions, a system may
cross the stability boundary during its motion. In such a situation, a
\textit{spontaneous contact mode transition} will occur. That is, the
system would switch from the $S_+S_+$ mode to a different type of
motion, despite the fact that the $S_+S_+$ mode remains consistent. As
we argued in Secs.~\ref{sec:3} and \ref{sec:4} such spontaneous
transitions are not possible in the single-contact case.
  
Another intriguing property of these sometimes unstable $S_+S_+$ regions is 
that their exact boundaries within a subclass appear to 
depend on the underlying compliant
contact model, even if the compliance parameter $\epsilon \to 0$.
Hence, unlike in the single-contact case, the stability of a
contact mode may be undecidable within rigid-body theory.

A closer look at Table~\ref{tab:classification} reveals an even more
surprising phenomenon. In subclasses 261 and 262, the $S_+S_+$ is the
unique consistent contact mode. These cases appear to be free from any
paradoxes, but nevertheless the stability analysis shows that the
$S_+S_+$ mode is robustly unstable. Robustness is meant in the sense
of uniform resolvability, as introduced within Question 3 in the 
Introduction. That is, 
instability emerges 
as $\epsilon \to 0$ 
for any form of contact compliance within the
framework of Sec.~\ref{sec:3.2}. As we shall see in the next subsection
(see e.g.~Fig.~\ref{fig:unstableslip}(A)), an IWC is likely to 
happen in these cases. 
 
%
%
%
%

This observation refutes an often assumed hypothesis within the rigid
body mechanics community that the failure of an object to slip
smoothly along contact surfaces is always a result of the \pain
paradox. This philosophy is embodied in an {\em a priori} rule 
known as Kilmister's principle of constraints \cite{Brogliato1999}: 
{\em "a unilateral
constraint must be verified with (bounded) forces each time it is
possible, and with impulses if and only if it is not possible with
forces"}.  The observation also questions the relevance of enumerating
every possible consistent contact modes, which is often assumed in a
hybrid systems or complementarity formation to be the key question in
the analysis of mechanical systems with frictional contacts, see
e.g.~\cite{Lotstedt,Pang}.  


\subsection{Impacts without collision} \label{sec:6.3}

Let us now consider the possibility of impact, specifically an IWC for
each of the cases enumerated in Table \ref{tab:classification}.
This is indicated
in  last three columns of the table using a notation we shall now explain.
A system with multiple contacts may undergo several types of impacts:
an impulsive contact force may occur at any non-empty subset of the
contact points, while the rest of the contacts lift off. In the
simplest case of two non-sticking contacts, we have three types of IWC:
impulsive force at one contact while the other lifts off, which we label
by $IF$ or $FI$ depending on which compact impacts; 
and simultaneous impact at both contacts, which we label $II$. 

Provided that we retain the assumption that the response 
of the object follows rigid body-theory, the $IF$ impact 
is consistent if the contact force at contact 1 seeks to push the 
contact point into the ground and simultaneously seeks to detach the
other point. That is we require $p_{11}<0$ and 
the other point $p_{21}>0$. Analogously, $FI$ may occur only 
if $p_{22}<0$ and $p_{12}>0$. 

The case of the $II$
impact is more subtle. It is a necessary condition that there must 
exists a combination of positive impulsive contact forces at the two contacts,
which accelerates both contact points towards the ground. In other
words, the cone spanned by vectors $p_1$ and $p_2$ should intersect
the fourth quadrant of $\Rset^2$. Nevertheless, this
condition can be shown to not be sufficient. To study the 
simultaneous dynamics of the two
contacts during a double-impact process in general requires a careful
analysis using a compliant contact model. 

Thus the notation in the final three columns of Table \ref{tab:classification}
is to indicate with an `x' where impacts of each type are possible.
For the case of $II$ impacts, we have indicated by `(x)'  where the necessary
conditions are satisfied. 

\begin{figure}
\begin{center}
\includegraphics[width=0.75\textwidth]{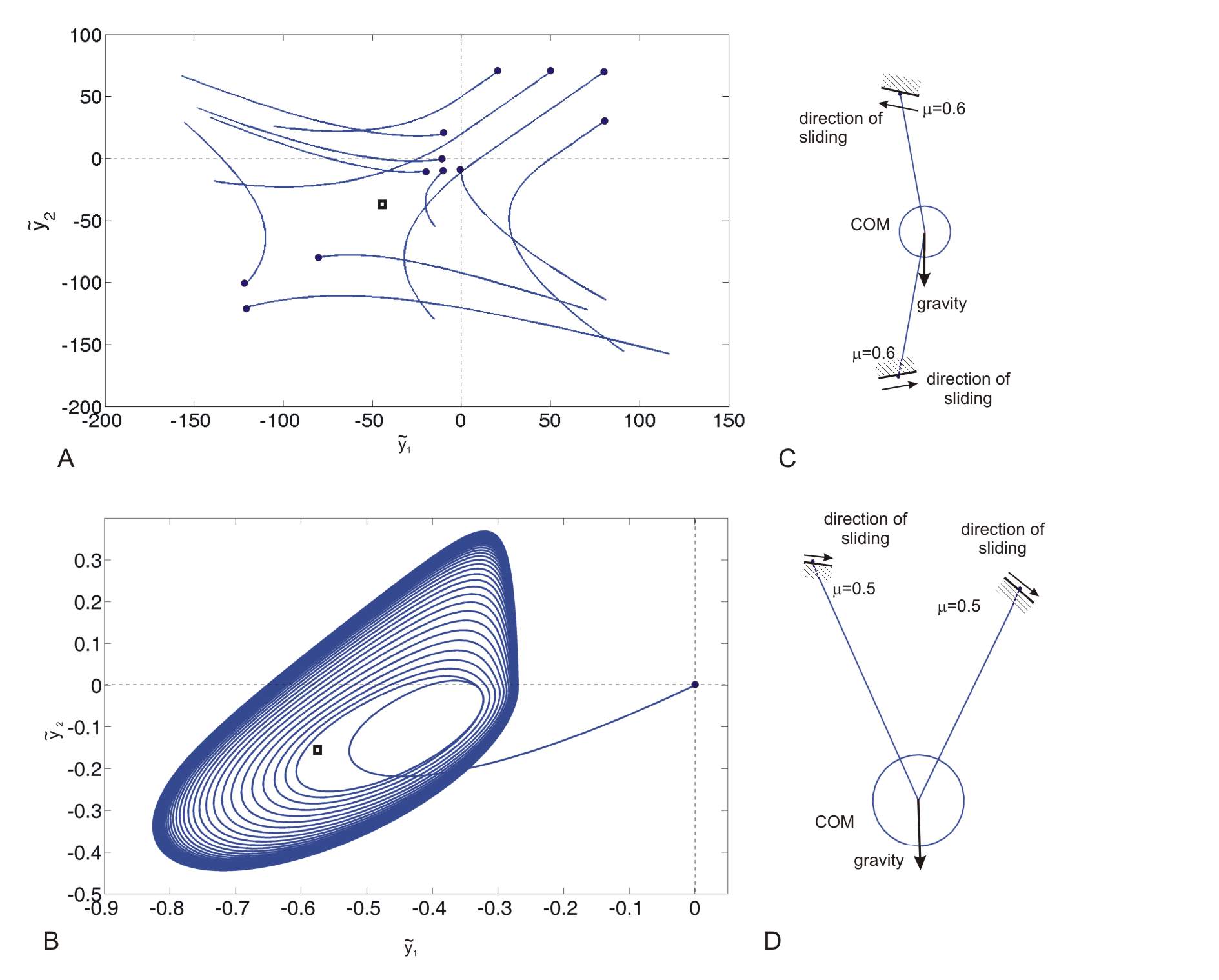}
\end{center}
\caption{Simulation of fast contact dynamics in examples of 
(A) class 261 and (B) class 142 using the contact model 
\eqref{eq:linearcontact}. See text for details. The corresponding
panels (C) and (D) show possible configurations that display the
dynamics in panels (A) and (B) respectively. Parameter values used are
(A) $a=[1\;1]^T$, $B=[[-0.055\; 0.055]^T\; [0.088\; -0.036]^T]$, inducing
  $\gamma_1=0.75\pi$; $\gamma_2=1.875\pi$; $\alpha=0.25\pi$, and 
(B) $a=[1\; 0.727]^T$,
$B=[[3.051\; 1.555]^T\; [-1.677\; -0.0375]^T]$, $k_1=1$; $k_2=3$;
$q_1=q_2=1.2$, inducing $\gamma_1=0.15\pi$; $\gamma_2=1.07\pi$; and
$\alpha=0.2\pi$. }
\label{fig:unstableslip}
\end{figure}

The results of what happens in a double impact $II$ 
can be shown to depend heavily on the exact form of the 
compliance model (satisfying the properties given  in Sec.~\ref{sec:4}), 
or equivalently of the impact
model.  The details will be presented
elsewhere \cite{Varkonyi_unp}. Computations reveal that there
exist cases in which every contact mode is either inconsistent or
unstable, and simultaneously, the three forms of IWC mentioned above
are impossible, see Fig.~\ref{fig:unstableslip}(B).
This observation points towards the incorrectness of
another common assumption: that \painx's paradox is perfectly
resolvable by considering IWCs. 

Figure \ref{fig:unstableslip}(A) shows a planar projection of the phase
portrait of the fast contact dynamics for a configuration within 
class 261 (where the $S_+S_+$ mode is the unique consistent contact mode, but it is unstable, see Sec. \ref{sec:6.2}). Several different initial conditions are depicted
in which the  initial time derivatives of
$\tilde{y}_i$ are 0. The contact model \eqref{eq:linearcontact} is
used in the simulation, with $k=q=-1$. For every possible initial condition, one of the two
contact penetrations, and
hence the normal force diverges to minus infinity. The other contact always 
lifts off. In the rigid limit $\epsilon\rightarrow 0$ this
behaviour corresponds to an IWC at
one of the two contacts, and lift off at the other, that is, either mode
$IF$ or mode $FI$. Note that the contact dynamics also has a consistent mode
$S_+ S_+$ in which both contacts slip, marked by a square in the figure, but
this is unstable. The stable manifold of this point in the fast dynamics
represents the dividing surface between two different types of IWC.

Figure \ref{fig:unstableslip}(B) shows an example within class 142 in
which there is the onset of some kind of micro-scale sprag-slip
oscillation. 
Here there is also an unstable 
$S_+S_+$ mode, again represented by a square. For the parameter values
depicted, every other contact mode is inconsistent and IWCs are
not possible either.  The initial condition for the trajectory
depicted is
$\tilde{y}_1=\tilde{y}_2=\dot{\tilde{y}}_1=\dot{\tilde{y}}_2=0$.  The
dynamics converges to an attractive limit cycle representing slipping
motion accompanied by fast micro-vibration, with repeated liftoff and
impact at the second contact point. In the limit $\epsilon \to 0$ this would 
correspond to a strange kind of sprag-slip limit cycle, for which the 
period is infinitesimally small. 

Panels C and D of Fig.~\ref{fig:unstableslip} show representations of
slipping rigid bodies possessing the values of matrix $B$ that correspond
to the cases shown in panels A and B, respectively. 
The bodies are under the influence of gravity, whose direction is
represented by the solid arrow through the centres of mass (COM). 
The orientation of the vector $b$ for each body, depends on the slip 
velocity, with the case illustrated corresponding to this velocity being
close to zero. The shape and dimension of each body and its point of 
contact are represented graphically by a circle that shows the 
radius of gyration centred at the COM, and two solid line sections 
terminating at the contact points. The orientations of the 
contact surfaces are represented by bold lines with hatched fills. The
values of the coefficient of friction is specified by a label at each contact. 

\subsection{Stability of equilibria}
\label{sec:6.4}

Stability of equilibrium configurations of rigid-body systems
with multiple unilateral contacts is of particular practical
importance in robotics, where one needs to be able to demonstrate
that grasping manoeuvres are robust.  In this context Pang and Trinkle
\cite{Pang} proposed various characterisations of stable equilibrium
configurations of systems with any number of contacts.  They refer to
a configuration as exhibiting the \textit{weak stability property} if the
$TT..T$ mode is consistent and the \textit{strong stability property} if
$TT..T$ mode is the only consistent contact mode which has $u=0$ for
all contacts.  However, they did not explore the consequences of these
notions of ``stability'' from the point of view of dynamical systems
theory. Howard and Kumar \cite{howard96}  demonstrated 
that the
\textit{strong stability} property of Pang and Trinkle is not necessary for
stability against small external forces (structural stability) 
for the case of a planar rigid
body with any number of contacts. An extension of this result to
multibody systems was given in \cite{Varkonyi15b}.

We found in Sec.~\ref{sec:3.3} that reverse chattering can occur in 
a system with a single contact, if $b<0$ and the effective restitution
coefficient $e>1$. We
also pointed out in Sec.~\ref{sec:3.4} 
the significance of avoiding reverse chatter in proving
Lyapunov stability of equilibria.
The case of multiple contacts is more complex because a sequence
of impacts may include impacts at any of the contact points in regular
(periodic) or irregular (chaotic) order. In some cases, simultaneous
impacts at multiple points are also to be expected.

Or and Rimon \cite{Or08b} were the first to point out that the key
to understanding Lyapunov stability of equilibria with 
unilateral frictional point contacts is to understand the 
possibility of reverse chattering motion of the system. 
They developed a sufficient condition for
Lyapunov stability of a planar rigid body with two point contacts
in the case of a specific impact law. A different sufficient condition using a
more realistic impact law was developed by V\'arkonyi \cite{Varkonyi12}, 
who also demonstrated that reverse chattering $e>1$ may occur even 
if impacts are
purely inelastic, i.e.~if $r_e=0$. More recently, 
V\'{a}rkonyi and Or found an almost exact
condition of stability for planar bodies with two contacts and $r_e=0$, 
the details of which will appear elsewhere 
\cite{Varkonyi-Or_unp}.

\begin{figure}
\begin{center}
\includegraphics[width=0.75\textwidth]{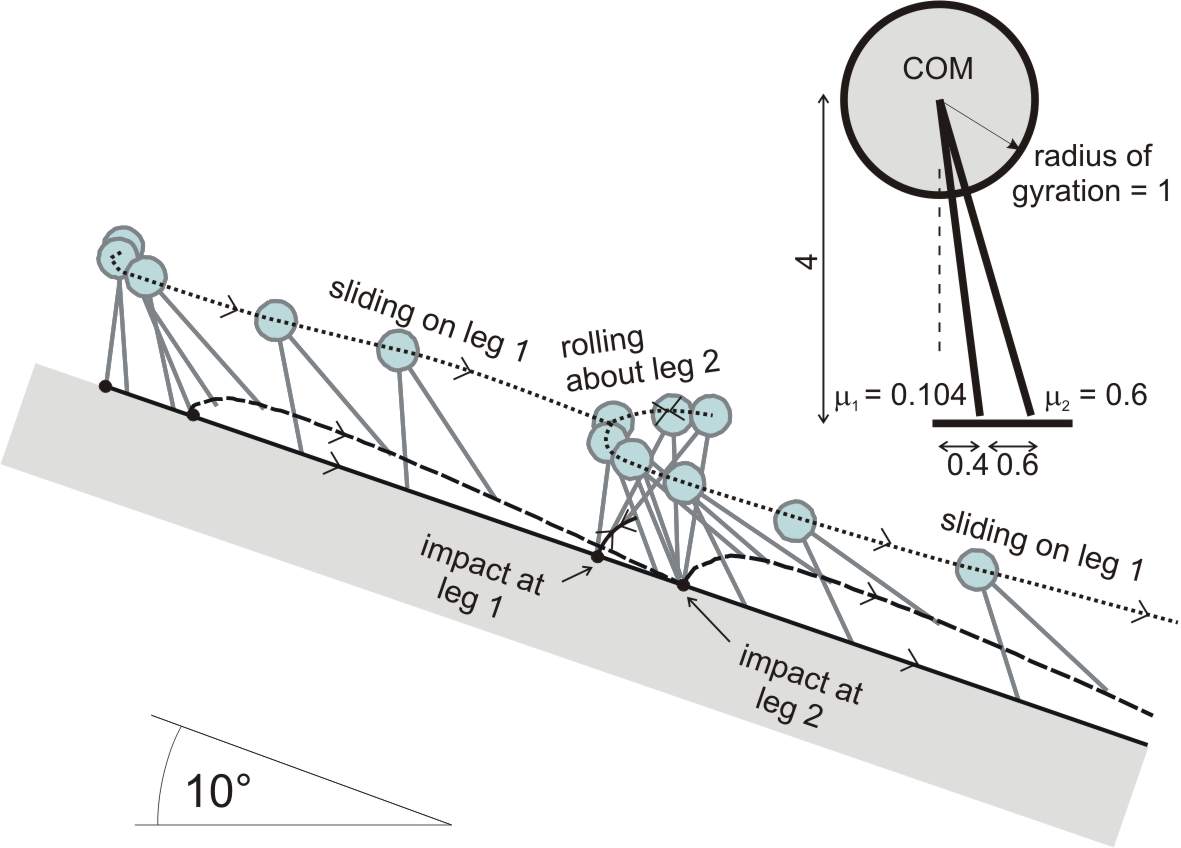}
\end{center}
\caption{(Adapted from \cite{Varkonyi12}). Simulation of the response of 
a heavy planar rigid body to a small perturbation of 
its equilibrium configuration on a perfectly inelastic ($r_e=0$) 
slope. See text for details.
\label{fig:instability}}
\end{figure}

Figure \ref{fig:instability} presents simulation results for a 
specific example of heavy planar
rigid body with two contact points that satisfies the strong stability property of \cite{Pang}, 
yet is not stable in the sense of Lyapunov. 
The shape of the body is represented by a circle and two lines 
as in Fig.~\ref{fig:unstableslip}, and for this simulation $r_e=0$.  
Initially, the object is resting on a slope of angle $10^\circ$, under
the influence of gravity. In this case, a static equilibrium equilibrium
in $TT$ mode can be shown to be consistent (calculation omitted). 
Nevertheless an infinitesimally small state-perturbation 
may set the object into divergent motion, causing it to tumble down
the slope as illustrated. First we find accelerating slip
motion on leg 1, in mode $S_+F$.  This is followed by a sticking impact at leg
2 followed by a rolling motion of that leg (i.e.~mode $FT$). This
mode continues until leg 1 hits the ground again. 
These four steps of motion are repeated again and again, while
the velocity of the object increases exponentially. The trajectories
of the two contact points and the COM are shown by solid, dashed, and
dotted curves, respectively in the figure. For more details, see
\cite{Varkonyi12}. 
It is interesting to note that if we had taken a lower coefficient of
friction at leg 1, specifically if $\mu_1<0.1$, then the initial equilibrium 
would become ambiguous. Thus at time $t=0$ there would be an additional
possibility for the motion, namely the behaviour displayed in the right-hand 
panels of Fig.~\ref{fig:ambiguity}.

If nothing else, this example illustrates that these considerations
of stability with multiple points of contact are not merely 
of academic interest, but can 
be easily manifest in purely passive mechanical systems. Another recent
practical illustration can be found in the work of Gamos and Or 
\cite{GamosOr}. There they compute many details of the dynamics of
bi-pedal passive walkers that can undergo stick-slip transitions,
including several cases of multi-stability with complex 
basins of attraction.

%
%
\section{Single-point-of-contact case in 3D}
\label{sec:7}

Everything discussed so has been for systems whose contact forces and
motions line in a plane. Much less is known
about the \pain paradox in situations that require a fully
three-dimensional analysis. For simplicity, we 
restrict attention to the single contact case.   
Our goal is to highlight similarities with and differences from 
the planar case. In what follows we make some general observations, but
only do precise calculations in the case of the three-dimensional analogue
of the CPP.

\subsection{Conditions for \pain paradoxes to occur}

For any spatial system with a single contact, equations of motion analogous to 
\eqref{eq:udynamics} and \eqref{eq:vdynamics} can be derived:
\begin{align}
\dot{u}& = a (q,\dot{q},t) + A(q,t) \lambda_{T}
+\lambda_{N} B( q,t ) ,  \label{eq:udynamics_3D} \\
\dot{v}& =b ( q,\dot{q},t ) +B^T( q,t ) \lambda_{T}
+\lambda_{N} C( q,t) ,  \label{eq:vdynamics_3D}
\end{align}  
where now the tangential velocity $u \in \Rset^2$ is a 2-vector, 
as are $a$, $B$, and $\lambda_T$, 
whereas $A$ becomes a $2\times 2$ matrix. In contrast, quantities involved
with normal directions --- $v$, $a$, $C$ and $\lambda_N$ --- remain scalars.

\begin{figure}
\begin{center}
\includegraphics[width=\textwidth]{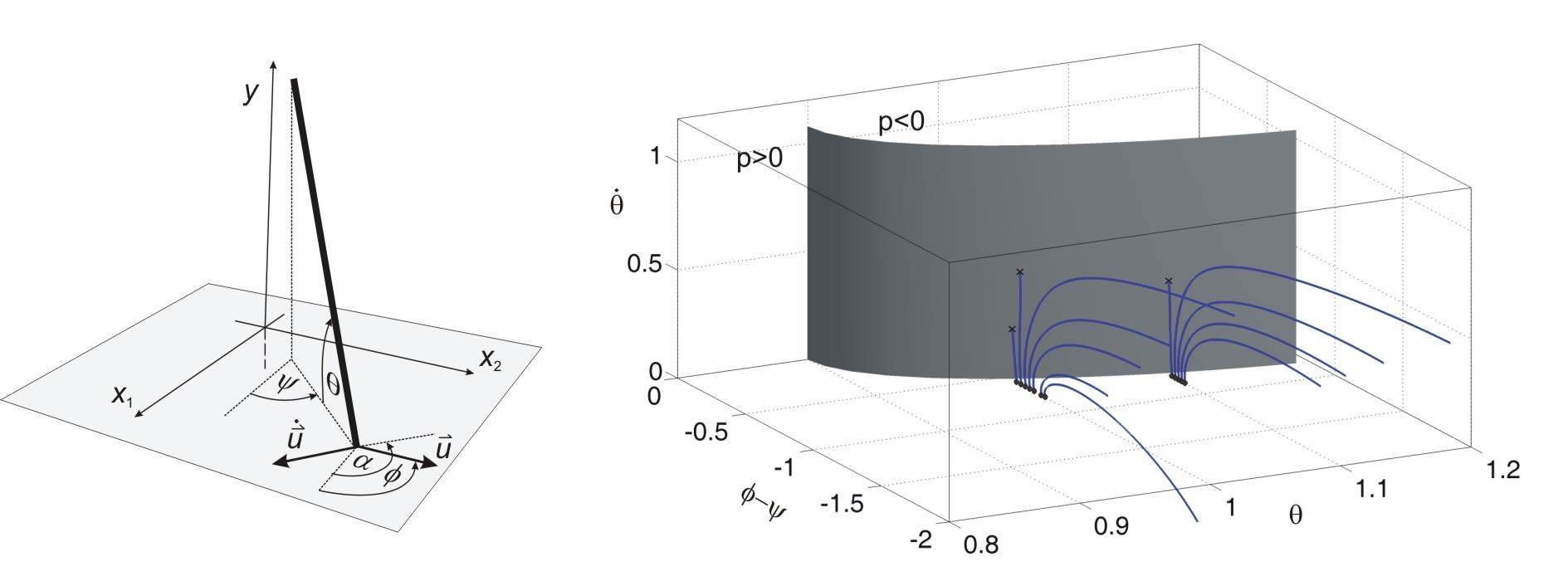} 
\end{center}
\caption{left: a slender, three-dimensional rod slipping 
on a horizontal surface. The coordinate axes $x_1$ and $x_2$ are horizontal, 
whereas axis $y$ is vertical. Here $u$ and $\dot{u}$ are the 
velocity and the acceleration of the contact point, respectively. 
Right: simulated trajectories of the slipping rod in a
three-dimensional projection of state space, with the boundary of the 
\pain region represented by the grey surface. Circles
denote initial points of the trajectories and x-shaped markers
indicate points of crossing $p=0$. Parameter values used are
$\mu=1$ and radius of gyration $=0.1 l$}

\label{fig:3Drod}
\end{figure}

A three-dimensional analogue of the Coulomb friction law can be written in 
the form
\begin{equation}
\lambda_T=-\mu\lambda_N \left[   \cos\phi \quad \sin\phi  \right]^T,
\label{eq:frictionlaw_3D}
\end{equation}
where $\phi$ is the angle of $u$ relative 
to the positive $x_1$ axis,
as in Fig.~\ref{fig:3Drod}. Equations \eqref{eq:vdynamics_3D}  
and \eqref{eq:frictionlaw_3D} imply
\begin{equation}
    \dot{v} = b+ (C-\mu B^T [ \cos\phi \quad \sin\phi ]^T )\lambda_N =0,
    \label{eq:vdotslip+_3D}
\end{equation}
which leads naturally to the definition of the three-dimensional
generalisation of the \pain parameter
\begin{equation}
    p(\phi):= C-\mu B^T [   \cos\phi \quad \sin\phi ]^T .
    \label{eq:p-3D}
    \end{equation}
Note that for motion in a plane with angle $\psi$ to the 
$(x_1,y)$ plane, as in Fig.~\ref{fig:3Drod}, 
sliding in $S_+$ mode would correspond to $\phi=\psi$,
and sliding in $S_-$ mode
to $\phi=\psi+\pi$ if $0<\theta<\pi/2$. 

Exactly as in the planar case, the free fall mode 
is consistent if there is no 
penetration into the contact surface, that is if $b>0$. Moreover the
slip mode $S$ is consistent if $\lambda_N>0$, 
which leads to the same condition $bp<0$. Hence, there is a unique
consistent mode (either free fall or slip) if $p>0$. In contrast, either both
modes are consistent or none of them is consistent if $p<0$.

As an example, consider the
three-dimensional analogue of the CPP, 
namely a slender, rod
with uniform cross-section contacting obliquely with
a frictional surface.  
Zhao \textit{et al} \cite{Zhao2008b} derived the 
specific equations of motion for such a rod, which when
projected onto tangential and normal directions 
lead \eqref{eq:udynamics_3D} and \eqref{eq:udynamics_3D} 
with 
\begin{align}
a & = - l
\left[ \begin{array}{c}
 \dot\psi^2\cos\psi\cos^3\theta + \dot\theta^2\cos\psi\cos\theta  \\
   \dot\psi^2\sin\psi\cos^3\theta + \dot\theta^2\sin\psi\cos\theta                \end{array}\right]
   \label{eq:a-3D}\\
b&=l\dot\psi^2\sin\theta+l\dot\theta^2\sin\theta-g  \label{eq:b-3D}  \\
A & =  m^{-1} 
\left[\begin{array}{cc}
1+3\sin^2\psi+3\cos^2\psi\sin^2\theta & -3\sin\psi\cos\psi\cos^2\theta  \\
-3\sin\psi\cos\psi\cos^2\theta & 1+3\cos^2\psi+3\sin^2\psi\sin^2\theta  
\end{array}\right]  \label{eq:A-3D} \\
B & =   3m^{-1}\sin\theta\cos\theta
\left[\begin{array}{c}
\cos\psi  \\ \sin\psi 
\end{array}\right]  \label{eq:B-3D} \\
C &= m^{-1}\left( 1+3\cos^2\theta\right).   \label{eq:C-3D}
\end{align}  
Here, $m$ is the mass and $l$ is the half-length of the rod, $\theta$
is the angle between the rod and the horizontal plane. 
The vectors $u$ and $\lambda_T$ are
expressed in a global $(x_1,x_2)$ reference frame, see Fig.~\ref{fig:3Drod}.

Let us consider the possibility of the \pain paradox 
for this problem.
With the help of \eqref{eq:b-3D},\eqref{eq:B-3D}, \eqref{eq:C-3D}, the
condition $p<0$ for the \pain paradox to occur can be expressed as
\begin{equation}
   \mu \cos(\phi-\psi) > \mu_p(\theta) = \frac{1+3\cos^2\theta}{3\cos\theta\sin\theta},
    \label{eq:painleveness-3D}
\end{equation}
which is identical to the condition for the planar rod
\eqref{eq:mumin} if $\phi=\psi$. 
Thus the \pain paradox for the 3D slender rod
cannot occur if $\mu\leq\mu_P(\theta)$ or if $\cos(\phi-\psi)\leq 0$. In the
case of $\mu\geq\mu_P(\theta)$ the paradox occurs in a range of
slip directions parametrised by 
$\phi-\psi \in (\phi_{\rm min},\phi_{\rm max})$,  where 
$\phi_{\rm max}=0$ if $\mu=\mu_P(\theta)$ and $\phi_{max}$ grows to $\pi/2$ if
$\mu\rightarrow\infty$.

It was also shown in \cite{Zhao2008b} that if the rod is in the
\pain regime with $b<0$, the non-existence paradox (inconsistency) 
can be resolved by
assuming an IWC. This assumption was also
justified by taking a compliant contact model similar in fashion to the one
introduced in Sec.~\ref{sec:4}. To obtain a more detailed picture of
the dynamics in the \pain regime, Shen \cite{Shen} considered a
elastic finite element model of the CPP, using a 3D rod of a
finite thickness with a rounded end. He was able to compare the
dynamics during planar motion of the system in its plane of symmetry
to the dynamics of an equivalent point contact model with a compliant
contact point. He also argued that both models show that the
continuation in the \pain regime can be understood in terms of an
IWC.

\subsection{Dynamic jam }

So far, we have not uncovered any behaviour of 3D systems 
that is qualitatively different from the planar case, in terms of 
consistent contact modes or the emergence of IWC.
However, when we consider Question 2 from the introduction we get a more
subtle answer. In particular, we found in 
Sec.~\ref{sec:5.1}, that it was 
not possible to enter the \pain 
region through the set $p^+=0$ while slipping, except
at the G-spot where $b$ is simultaneously zero. 
As we shall now show, this need not be the case for 3D systems. 

A fundamental difference that distinguishes the 2D and 3D cases is 
that for planar systems, the \pain parameters $p^+$
and $p^-$ depend on position variables $q$ and coefficient of friction
only. Whereas, note from \eqref{eq:p-3D} that 
$p(\phi)$ in general
depends additionally on the generalised velocities $\dot{q}$, through
the definition of the angle $\phi$.  Note that the 
angle $\psi$ is a cyclic coordinate of the slipping rod, hence we
may assume $\psi=0$ without loss of generality. Then,
\eqref{eq:udynamics_3D} can be expressed with the help of
\eqref{eq:a-3D}, \eqref{eq:A-3D}, \eqref{eq:B-3D} and
\eqref{eq:frictionlaw_3D} as
\begin{equation}
m\dot{u}=-l\cos\theta\left[
\begin{array}{c}  
\dot\psi^2\cos^2\theta+\dot\theta^2 \\  
0 
\end{array} \right] -
\mu\lambda_N
\left[ \begin{array}{cc}
1+3\sin^2\theta & 0\\
0 & 4 
\end{array} \right]
\left[ \begin{array}{c}
\cos\phi \\
\sin\phi
\end{array} \right]
+3\sin\theta\cos\theta\lambda_N
\left[\begin{array} {c}
1\\
0
\end{array}\right]
\label{eq:tangentialdyn-3D}
\end{equation}

Assume now that this system is very close to but outside of the \pain
regime, such that $0<p\ll1$. Furthermore, the free acceleration is such
that the rod remains in contact ($b<0$), 
which is the case, for example, if the rod has
sufficiently small angular velocity. According to
\eqref{eq:vdotslip+_3D} we find that 
$\lambda_N=-b/p \gg 1$. Thus, $O(1)$ terms in
\eqref{eq:tangentialdyn-3D} become negligible in comparison with
$O(\lambda_N)$ terms, yielding
\begin{align}
-m\lambda_N^{-1}\cdot \dot u &  = 
\mu
\left[ \begin{array}{cc}
1+3\sin^2\theta & 0\\
0 & 4 
\end{array} \right]
\left[ \begin{array}{c}
\cos\phi \\
\sin\phi
\end{array} \right]
-3\sin\theta\cos\theta
\left[\begin{array} {c}
1\\
0
\end{array}\right]
+O(\lambda_N^{-1}) \nonumber \\
& =  
\mu
\left[ \begin{array}{cc}
(1+3\sin^2\theta)\mu-3\sin\theta\cos\theta\cos^{-1}\phi & 0\\
0 & 4 \mu
\end{array} \right]
\left[ \begin{array}{c}
\cos\phi \\
\sin\phi
\end{array} \right]
+O(\lambda_N^{-1}) .
\label{eq:tangentialdyn2-3D}
\end{align}

The upper left element of the diagonal matrix in
\label{eq:tangentialdyn2-3D} 
is smaller than the lower right one because
\begin{equation}
(1+3\sin^2\theta)\mu-3\sin\theta\cos\theta\cos^{-1}\phi  \leq (1+3)\mu-0=4\mu.
\end{equation}
This means that the absolute value of the angle of the vector
that $-\dot u$ makes relative to the positive $x_1$ axis
is larger than the absolute value of the angle
$\phi$ of $u$.
Therefore, $\phi$ approaches 0 over time 
in such a way that $\dot\phi\cdot\sin\phi < 0$. Meanwhile the time
derivative of the \pain parameter $p$ is
\eqref{eq:vdotslip+_3D}
\begin{equation*}
   \dot p=\dot C-  \frac{d}{dt} \left(\mu B^T\right)    \left[  \cos\phi \quad \sin\phi  \right]^T
   - \mu B^T\frac{d}{dt}{\left[   \cos\phi \quad \sin\phi  \right]^T}
\end{equation*}
in which all terms are $O(1)$ except for the acceleration-like term
$\frac{d}{dt}\left[ \cos\phi \quad \sin\phi \right]$, which is
$O(\lambda_N)$. Neglecting the $O(1)$ terms and substituting $\psi=0$ leads
to
 \begin{align*}
 \dot p & =  - \mu B^T\frac{d}{dt}{\left[   
\cos\phi \quad \sin\phi  \right]^T} + O(1) \\
 &= 3\mu m^{-1}\sin\theta\cos\theta\sin\phi\dot\phi + O(1),
 \end{align*}
which is negative because $\sin\phi\dot\phi<0$ and all other terms are
positive. Thus we arrive to the conclusion that $p$ becomes negative,
and the system crosses the boundary of the \pain regime transversely
away from the G-spot.  Our conclusion is confirmed by numerical
simulation (Fig.~\ref{fig:3Drod}, right panel). What happens beyond
this crossing is an open question.

\section{Discussion}
\label{sec:8}

This article represents an attempt to survey the state of knowledge on
the range complex phenomena in rigid body mechanics 
that have been grouped together under the
epithet of the \pain paradox. We have drawn particular attention to recent
results by the authors and their collaborators, while trying to establish some
general notation and terminology that we hope will help to organise the 
previous literature and stimulate other researchers to investigate this
fascinating topic. Let us close with some philosophical
remarks and a brief summary of open problems. 

%

\subsection{Philosophical remarks}

Far from being resolved, the Painlev\'{e} paradox on the inconsistency of
rigid body mechanics with frictional contact, first posed over 120 years ago, 
would seem to retain many open facets. Even for the highly restrictive 
situation of a planar configuration with a single point contact, we
have argued that there remain situations that are undecidable within
the framework of rigid body mechanics. In particular, slender bodies
such as the classical \pain problem of the falling rod reach a point
of dynamic jam, from which it is not currently clear whether there is
a unique forward continuation without including additional physics. 
In three dimensions, we have shown that
things are even worse, because the rod can enter the unstable \pain
region transversely. Here, the results in Table \ref{tab:summary2}
suggest there is indeterminacy because the rod could lift-off or take
an impact. Also, even outside the \pain region, we argued in
Sec.~\ref{sec:5} that reverse chatter can be
triggered at a transition into stick.  If reverse chatter occurs then
there is infinite degeneracy between a continuum of different
phases. Nevertheless, for the single contact case in 2D we showed that
inclusion of an impact law and an analysis using contact regularisation
enables resolution of all inconsistent cases, but indeterminacy remains.

If one allows more than one contact point, then we indicated in 
Sec.~\ref{sec:6} that
there are many other cases of indeterminacy, and we have not yet been able
to resolve all possible cases of inconsistency, even under the
simplifying assumption that no contact is in stick.  
We have also provided evidence, for
example through the micro-scale chatter-like behaviour of Fig.~\ref{fig:unstableslip}, that there may be far more complex
possibilities for reverse or forward chatter-like behaviours 
in the case of two or more contacts. 

One thing that our analysis does point to though is the insufficiency
of Kilmister's principle applied to rigid bodies with multiple
contacts. That has led to a common approach in multibody mechanics to
seek sufficient conditions under which there is a unique consistent
contact mode, thus avoiding any \pain paradoxes.  
We argue that this approach is misguided, because we already
know from the single point of contact case that there are easily
reachable configurations that are inevitably undecidable and which
cannot be resolved without breaking the assumptions of rigid body
mechanics.  We also showed a specific case in Sec.~\ref{sec:6} where
there is a unique consistent contact mode $S_+ S_+$, but this is not
normally stable and instead an IWC must occur. Rather than seek
sufficient conditions to sweep the paradox under the carpet, as it
were, we have been motivated instead by the three separate questions
posed in the Introduction. In particular we have attempted to separate
the questions of: which configurations are consistent or determinate;
how one may undergo a transition into an indeterminate state; and
whether questions of consistency or indeterminacy can be resolved by
consideration of impact and contact regularisation.

Contact regularisation concerns the introduction of finite
stiffnesses and in Sec.~\ref{sec:4} we considered only the simplest
kind, namely introduction of a stiffness in the normal degree of freedom. 
 In one sense such normal regularisation can resolve all of the
consequences of the \pain paradoxes, if one allows the stiffness to
remain finite.  Even in this case though, simple nonsmooth friction
laws in the tangent direction lead to Filippov systems which can
themselves exhibit indeterminacy; see \cite{Jeffrey} for a mechanical
example.  The test we have used for genuine determinacy associated
with the \pain paradox is what we have termed {\em uniform resolvability},
namely that normal contact regularisation leads to unique limiting
behaviour in the infinite stiffness limit. If such a unique limit does
not occur, such as in reverse chatter, then these cases we deem to be
genuinely indeterminate. Of course, there are many other possible
approaches to regularisation, and we mentioned already the approach
of Szalai \cite{Szalai1,Szalai2} that appears promising. 

We should also stress the fundamental point made in the Introduction that
the \pain paradox is not about the real world, but an indication of
the failure of rigid-body mechanics.
Typically any friction or impact law is a gross approximation to
tribological processes that occur at the microscale.
Thus, the existence of undecidability (either inconsistency or
indeterminacy) within a rigid body framework points to situations for
which there will be extreme sensitivity with respect to initial
conditions in the real mechanics.

An overarching conclusion then is that, for certain configurations at
least, a purely rational mechanics approach to rigid body dynamics
with impact and friction is fundamentally ill-posed.  To put it
somewhat pejoratively, one might say that \textit{there is nothing
  ``rational'' about rational mechanics} at least for rigid bodies.
There is a strong tradition going back to Newton, Euler and many
others in trying to discover the ``laws of mechanics'' and showing
that they are consistent. This body of literature has become known as
rational mechanics; and indeed the \textit{Archives for Rational
  Mechanics and Analysis} is one of the most prestigious scientific
journals. The hope is that one day, by posing sufficiently many
constitutive laws one can come up with self-consistent continuum
theories for fluids, elastic solids, elasto-plastic materials, etc.
One of the simplest such continuum theories is that of rigid body
mechanics, which we typically teach to our students through an
axiomatic approach. We talk about the Coulomb or Stribeck
\textit{laws} of friction, rather than models of friction.
Nevertheless we have shown that with even the simplest consistent
friction and impact laws for contact with friction lead to
fundamentally undecidable cases, and the search for a grand unifying
law that resolves all cases is probably futile.

Rather like Russell \& Whitehead's attempt in \textit{Principia
  Mathematica} \cite{Russell} to systematise arithmetic and, by
implication, the whole of mathematics (see, for example
\cite{Hofstadter} for a popular account), we could argue that any
mission with the aim rationalising rigid body mechanics through axioms
and laws is doomed to failure.  Rather like the way G\"{o}del's
Incompleteness Theorem \cite{Godel} showed the impossibility of
Russell and Whithead's goal, in our view, the inconsistency and
indeterminacy that is inherent in the \pain paradox suggests that a
complete, self-consistent, theory of rigid body mechanics is
impossible. It is also tempting to draw the analogy between
G\"{o}del's use of paradox at the heart of his proof. If G\"{o}del's
work has taught us anything it is that no mathematical theory of
sufficient complexity to describe the real work is ever complete;
undecidability paradoxes are inevitable.

\subsection{Open problems and perspectives}

In Sec.~\ref{sec:2.3} we reviewed some of the configurations that have
been shown to exhibit \pain paradox phenomena. Many of the practical
implications of the theory to these and related mechanisms remain
unexplored. For example, is it possible to explain the transition to
sprag-slip oscillation in brake-disc-like systems through a reverse
chatter instability?  Can such instabilities be used to explain the
mechanism for drawing dashed lines on a blackboard?  Another practical
area which seems largely unexplored is in rotor machinery with
contact, where there are large couplings between tangential and normal
motion due to gyroscopic forces. Even ignoring possible sprag-slip
behaviour, rotordyanmics with contact can lead to rich dynamics, see
e.g.~\cite{Zilli} and references therein.

With two or more contacts we have pointed in Sec.~\ref{sec:6} to many
new phenomena, including IWCs that can involve one or more contacts
and possible micro-chatter.  To use a deliberate pun, we have {\em
  only just scratched the surface} for planar bodies with two contacts
and there are many cases that remain unexplored. In three
dimensions, and for more than two contact points, the conditions that
lead to reverse chatter remain completely unexplored. Elucidating when
reverse chatter can occur is an important question for applications,
for example in the design of grippers and manipulators in robotics.
We we have shown, the question of which mode of motion is observed
cannot be solved simply by looking at which modes are
feasible. Instead, we need to consider Lyapunov stability, which, using
the method of contact regularisation, can lead to an understanding of
where IWC and/or reverse chatter is either possible or inevitable.

We also pointed to recent work \cite{GamosOr} that analyse how \pain
paradoxes can occur in simple models of legged locomotion. There is a
growing body of work in biomechanics that uses ideas from dynamical
systems and bifurcation theory to understand human gate stability,
since the pioneering work of Alexander \cite{Alexander}, see
e.g.~\cite{Seyfarth} and references therein for recent results.  As
yet though, the complexities associated with the \pain
paradox seems largely unexplored in that literature.

Another practical question is if the extreme sensitivity associated
with the \pain paradox is unavoidable, how can we design
control laws to ameliorate its effects? The state of the art in
control of such systems seems to be to use the framework of
complementarity or hybrid systems, see for example the work of Heemels
and co-workers \cite{Heemels}.  However, as we have shown, a complete
understanding of the dynamics requires consideration of contact
regularisation and impact, which go beyond the complementarity
view.

There are many theoretical questions that remain yet to be analysed to
determine consistency and stability.  As far as we are aware the
simplest cases yet to be analysed are situations where there are two sticking
contacts in 2D, or where there is just one sticking contact in 3D.  In
general we are long way from a complete list of types of dynamical
behaviour, and it is not yet known whether and how the non-existence
paradox can be resolved in all cases.  Also, characterisation of
bifurcations and transitions in multi-contact systems remains
completely unexplored. Implications of \pain paradoxes for the much
more tricky situations of line or regional contact seem to remain a
long way off. Even seemingly simple questions for planar single
contact systems remain unsolved, such as what happens beyond the
$G$-spot, and whether it is ever possible to cross
into $p<0$ other than at this point for systems that have more than
three-degrees of freedom. 

In summary, far from being resolved, there seems to be a rich array of
open questions, both theoretical and practical, that arise from the
indeterminacy, multistability and instability associated with the
classical \pain paradox. If nothing else we hope this survey will
serve to stimulate fresh efforts by the next generation of
researchers. It would seem to us that there could easily be room for
another 120 years of fruitful research on these fascinating phenomena.

\bibliographystyle{plain}
\bibliography{painleve}

\end{document}